\documentclass[a4paper, amsfonts, amssymb, amsmath, reprint, showkeys, nofootinbib, twoside,aps,prc, preprintnumbers]{revtex4-2}

\usepackage[english]{babel}
\usepackage[utf8]{inputenc}
\usepackage[colorinlistoftodos, color=green!40, prependcaption]{todonotes}
\usepackage{xcolor}
\usepackage{xspace}
\usepackage{lineno}
\usepackage{amssymb}
\usepackage{url}
\usepackage{ulem}

\usepackage{amsthm}
\usepackage{mathtools}
\usepackage{physics}
\usepackage{graphicx}
\usepackage[left=23mm,right=13mm,top=35mm,columnsep=15pt]{geometry} 
\usepackage{adjustbox}
\usepackage{placeins}
\usepackage[T1]{fontenc}
\usepackage{csquotes}
\usepackage{comment}
\usepackage[caption=false]{subfig}

\usepackage[Symbolsmallscale]{upgreek}
\usepackage[pdftex, pdftitle={Article}, pdfauthor={Author}]{hyperref} 
\usepackage{microtype} 

\newcommand{\La}{{\Lambda}}
\newcommand{\Si}{{\Sigma}}

\begin{document}
\title{Properties of hyperons in nuclear matter from chiral 
hyperon-nucleon interactions \\at next-to-next-to-leading order}

\author{Asanosuke Jinno}
\email{jinno.asanosuke.36w@st.kyoto-u.ac.jp}
\affiliation{Department of Physics, Faculty of Science, Kyoto University, Kyoto 606-8502, Japan}
\author{Johann Haidenbauer}%
\email{j.haidenbauer@fz-juelich.de}
\affiliation{Institute for Advanced Simulation (IAS-4), 
Forschungszentrum J\"ulich, D-52425 J\"ulich, Germany}
\author{Ulf-G. Mei{\ss}ner}%
\email{meissner@hiskp.uni-bonn.de}
\affiliation{Helmholtz-Institut~f\"{u}r~Strahlen-~und~Kernphysik~and~Bethe~Center~for~Theoretical~Physics,
~Universit\"{a}t~Bonn,~D-53115~Bonn,~Germany}
\affiliation{Institute for Advanced Simulation (IAS-4), 
Forschungszentrum J\"ulich, D-52425 J\"ulich, Germany}
\affiliation{Peng Huanwu Collaborative Center for Research and Education,
International Institute for Interdisciplinary and Frontiers, Beihang
University, Beijing 100191, China}

\preprint{KUNS-3074}

\begin{abstract}

The $\Lambda$ and $\Sigma$ single-particle potentials in infinite nuclear matter
are analyzed within a recently established
chiral hyperon-nucleon ($YN$) interaction up to N$^2$LO 
in combination
with a nucleon-nucleon interaction derived in the same scheme.
The self-consistent Brueckner-Hartree-Fock method with the continuous choice of the
single-particle potential is employed. It is found that the $\La$ single-particle
potential is comparable to the results achieved with the NLO $YN$ interaction from 2019.
The resulting $\Si$ potential becomes more attractive compared to the previous NLO results
due to the constraint from the recent $\Sigma N$ differential
cross section data measured in the J-PARC E40 experiment.
An estimate of the theoretical uncertainty of the single-particle potentials
is provided in terms of the truncation error in the chiral expansion.

\end{abstract}

\maketitle

\section{Introduction}

Due to the limited amount of direct experimental information on the
interaction of the hyperons $\La$ and $\Si$ with nucleons,
see Ref.~\cite{Haidenbauer:2023qhf} for a recent overview, 
it is essential to consider also their interaction in few- and
many-body systems as a test and as an additional source of information. 
This aim can be achieved within so-called \textit{ab initio} calculations
where the elementary hyperon-nucleon ($YN$) interaction is used as input
in Faddeev-Yakubovsky equations 
\cite{Miyagawa:1993rd,Nogga:2001ef,Nogga:2013pwa,Haidenbauer:2021wld}
or in the no-core shell model
\cite{Wirth:2014apa,Wirth:2017lso,Le:2020zdu,Le:2022ikc,Le:2024aox}
in order to evaluate the properties of light $\La$ hypernuclei. 
A similar strategy
is also followed in studies within nuclear lattice effective field theory
\cite{Hildenbrand:2024ypw}.
For heavier hypernuclei traditionally microscopic calculations have been 
performed with an effective $YN$ interaction, which is generated from the 
$G$ matrix, i.e., from the solution of the Bethe-Goldstone equation in
infinite nuclear matter for the underlying bare $YN$ potential
\cite{Yamamoto:1994tc,Hiyama:1997ub,Vidana:1998ed,
Fujiwara:2006fj,Vidana:2016ayd,Haidenbauer:2019thx}. 

The $YN$ $G$ matrix is also a key element in microscopic studies of compact 
objects like neutron stars 
\cite{Haidenbauer:2016vfq,Gerstung:2020ktv,Logoteta:2019utx,
Vidana:2024ngv} in the context of the so-called ``hyperon
puzzle``; see, e.g., Refs.~\cite{Glendenning:1991es,Knorren:1995ds,Balberg:1997yw,
Nishizaki:2002ih,Weissenborn:2011ut,Togashi:2016fky,Fortin:2017cvt,Tong:2024jvs,Fujimoto:2024doc}
and recent reviews~\cite{Tolos:2020aln,Burgio:2021vgk,SchaffnerBielichBook,Vidana:2022tlx}.
Here, ``puzzle'' refers to the quest to reconcile
the seemingly contradictory observations that in calculations
with standard $YN$ interactions
$\La$'s (and possibly other hyperons) appear with increasing density, 
thereby causing a softening of the equation of state (EOS), 
while the observed mass-radius relation and the maximum mass of 
neutron stars can be explained only with a stiff EOS.
Clearly, while for the application in finite nuclear matter
the $G$-matrix is usually used around nuclear matter saturation 
density, $\rho_0=0.16$~fm$^{-3}$, or below, with regard to neutron stars 
the relevant region extends very well up to ($6$-$7)\rho_0$.

For many years phenomenological $YN$ potentials have been used
as a starting point for such microscopic calculations.
However, this has changed in recent years with the arrival
of $YN$ potentials derived within 
chiral effective field theory (EFT). That approach with an inherent power
counting, initially suggested by Weinberg for application to the 
nucleon-nucleon ($NN$) interaction,   
allows one to improve the results order by order and, equally importantly, enables also an estimation of the theoretical uncertainty. 
This is not only valid for the elementary $YN$ interaction but applies also for few- and many-body studies where 
chiral $YN$ interactions are employed. 

In the present work we investigate the in-medium properties of the 
$\La$ and $\Si$ hyperons employing chiral $YN$ interactions. The study
is performed within the conventional Brueckner theory 
\cite{Brueckner:1954zz,Brueckner:1955zzb,Day:1967zza} at first order
in the hole-line expansion, i.e., the so-called Brueckner-Hartree-Fock
(BHF) approximation. The starting point is the new $YN$ potential 
published recently by the J\"ulich-Bonn group~\cite{Haidenbauer:2023qhf}, 
based on SU(3) chiral effective field theory up to next-to-next-to-leading 
order (N$^2$LO).
Besides extending the chiral order with respect to our earlier 
next-to-leading order (NLO) $YN$ potentials from 2013 (NLO13) \cite{Haidenbauer:2013oca} 
and 2019 (NLO19) \cite{Haidenbauer:2019boi}, 
the new $YN$ interaction builds also on a novel regularization scheme, 
the so-called semilocal momentum space regularization,
which has been shown to work rather well in the 
nucleon-nucleon ($NN$) sector \cite{Reinert:2017usi}. Furthermore, the 
potential incorporates new constraints from 
$\Si^-p$ and $\Si^+p$ differential cross sections measured in the 
J-PARC E40 experiment 
\cite{J-PARCE40:2021bgw,J-PARCE40:2021qxa,J-PARCE40:2022nvq}.
The new semilocal momentum space regularized 
(SMS) interactions are also the first $YN$ potentials where the 
separation energies of light $\La$ hypernuclei can be rather well
described \cite{Le:2024rkd}.

Our study has two goals. One of them is to provide a general overview of the 
in-medium properties that result from the new SMS $YN$ potentials. The other 
one, and maybe the more important one, is to explore the convergence pattern 
of the chiral expansion and the 
regulator dependence of the properties of hyperons in 
nuclear matter. Specifically, we
provide a first uncertainty estimate for the $\La$ and
$\Si$ single-particle potentials, following corresponding
efforts for nucleons in infinite nuclear matter
\cite{Hu:2016nkw,Hu:2019zwa}. 
So far in nuclear matter calculations with chiral $YN$ potentials 
only the cutoff dependence has been used to estimate the
uncertainty \cite{Haidenbauer:2014uua,Petschauer:2015nea,Mihaylov:2023ahn,Zheng:2025sol}.

An additional aspect we want to address is 
the influence of $YN$ $P$ waves on the matter properties.
Since experimental information on $\La p$ differential 
cross sections at low momenta is rather limited
\cite{Sechi-Zorn:pLambda,Alexander:pLambda},
the $YN$ interaction in the $P$ waves cannot be determined
reliably. 
In the case of NLO13 (NLO19) 
\cite{Haidenbauer:2013oca,Haidenbauer:2019boi}
the smallness of the $\La p$ cross section above the $\Si N$ threshold, 
i.e., at around $p_{lab} \approx 800$~MeV/$c$, was used as a guideline to
fix the strength of the $P$-wave interactions, together with
constraints from the corresponding $NN$ $P$-wave interactions
provided by the underlying SU(3) flavor symmetry. 
Agreement with the $\La p$ data could be only achieved if the 
individual contributions from all the $P$ waves 
($^1P_1$, $^3P_0$, $^3P_1$, $^3P_2$) 
are kept small, which means in turn that the corresponding
phase shifts have to be small. 
As already mentioned,
in the SMS $YN$ potential $\Sigma^-p$ and $\Sigma^+p$
differential cross sections, measured at J-PARC
\cite{J-PARCE40:2021bgw,J-PARCE40:2021qxa,J-PARCE40:2022nvq},
have been used to constrain the $P$ waves. However, also in this
case a unique determination of all $P$-wave interactions was not possible \cite{Haidenbauer:2023qhf}. 

We restrict our study to densities up to twice the 
nuclear matter saturation density $\rho_0 = 0.16$~fm$^{-3}$, 
i.e., to Fermi momenta below $k_F=2.15$~fm$^{-1}$
($\approx 400$~MeV) in pure neutron matter (PNM). 
This limit on the density is suggested by the breakdown scale of
the chiral expansion \cite{Epelbaum:2014efa,Epelbaum:2014sza}
and by the cutoff values
applied in the regulators of the chiral potentials
\cite{Haidenbauer:2013oca,Haidenbauer:2019boi,Haidenbauer:2023qhf}, which 
are both in the order of $500\pm 100$~MeV. 
See also the corresponding comments in Refs.~\cite{Hu:2016nkw,Kohno:2018gby}. 
Moreover, since the presence of the Pauli operator in the $G$-matrix
equation suppresses the
contributions from lower momenta in nuclear matter calculations 
one is automatically more sensitive to the high-momentum 
part and that means to regulator artifacts.
Extrapolations to higher density are, of course, always
possible, but one should keep in mind that they are 
predominantly of phenomenological nature.

The paper is structured in the following way.
Section~\ref{sec:theory} briefly introduces the baryon-baryon
interactions employed in this work and then
the Brueckner theory.
In Sec.~\ref{sec:results} we provide results for the single-particle
potentials of the $\La$ and the $\Si$ in symmetric nuclear matter (SNM)
and in PNM. We focus on the results at 
nuclear matter saturation density, but we also discuss 
the density and the momentum dependence of the single-particle potentials. 
In Sec.~\ref{sec:uncertaity} we perform an estimate of the theoretical 
uncertainty of our results for the hyperon single-particle potentials
due to the truncation in the chiral expansion, the first of its kind.
The paper ends with a brief summary and an outlook.

\section{Theoretical components}
\label{sec:theory}

\subsection{Hyperon-nucleon and nucleon-nucleon interactions}
\label{sec:YNandNN}

Recently the J\"ulich-Bonn group has established a $YN$ potential 
for the strangeness $S=-1$ sector ($\Lambda N$, $\Sigma N$) 
up to N$^2$LO in the chiral expansion
\cite{Haidenbauer:2023qhf}. The main focus of the present work is to 
explore the in-medium properties of $\La$ and $\Si$ that follow
from this interaction. Below we briefly summarize the main features
and merits of this interaction. For a detailed description we refer the reader to the original publication. 

\begin{figure}[tb]
    \centering
        \includegraphics[width=0.95\linewidth]{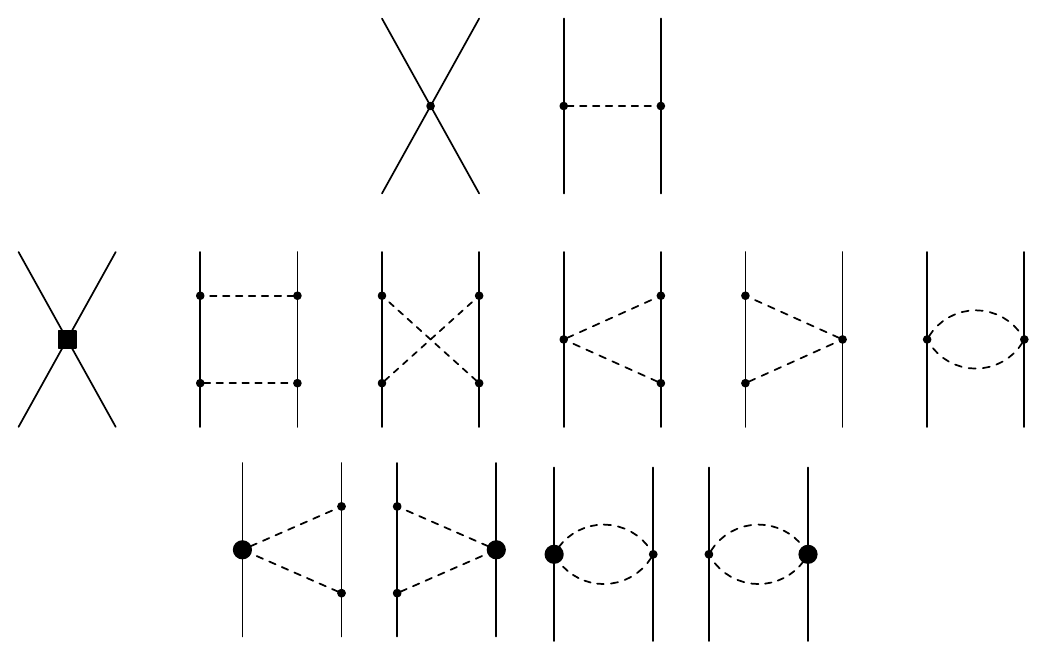}
    \caption{
        Diagrams contributing to the $YN$ potential at LO (top), NLO (center), and N$^2$LO (bottom) in chiral EFT. Solid and dashed lines denote 
        octet baryons and pseudoscalar mesons, respectively.
    }
    \label{fig:chEFT}
\end{figure}

The $YN$ interaction is derived within 
SU(3) chiral effective field theory with the Weinberg power counting applied 
to the potential. At LO one-meson exchange diagrams and nonderivative 
four-baryon contact terms contribute to the potential; see 
Fig.~\ref{fig:chEFT} (top). At NLO additional contact terms and (irreducible)
two-meson exchange diagrams at the one-loop level contribute (center). 
Finally, at N$^2$LO two-meson exchange contributions involving the
subleading meson-baryon Lagrangian arise (bottom).
The contributions from pseudoscalar-meson exchanges ($\pi$, $\eta$,
$K$) are, in principle, fixed by the assumed SU(3) flavor symmetry. 
The symmetry provides relations between the coupling constants at the
various meson-baryon-baryon vertices so that they all can be written in terms
of the $F$ and $D$ couplings of the underlying meson-baryon interaction
Lagrangian \cite{Haidenbauer:2013oca,Polinder:2006eq}.
$F$ and $D$ satisfy the relation $F + D = g_A \approx 1.27$, where $g_A$ is 
the axial-vector strength measured in neutron $\beta$ decay.
In contrast the contact terms represent the unresolved short-distance 
dynamics and the associated low-energy constants (LECs) are essentially
free parameters. In practice, they are fitted to low-energy $YN$ scattering 
data and additionally constrained by the hypertriton ($^3_\La$H) binding 
energy. At the order considered the contact terms yield contributions to 
the $S$ and $P$ waves of the potential. 

Although SU(3) flavor symmetry is imposed for constructing the 
interaction, the explicit SU(3) symmetry breaking by the 
physical masses of the pseudoscalar mesons ($\pi$, $K$, $\eta$) is taken
into account when evaluating the potential. The mass differences induce 
also a possible SU(3) breaking in the leading-order contact terms 
\cite{Haidenbauer:2013oca,Petschauer:2013uua}, which is likewise considered 
in the $YN$ potential of Ref.~\cite{Haidenbauer:2023qhf}. 

When solving the scattering equations for chiral potentials a regulator
has to be introduced in order to remove high-momentum components~\cite{Epelbaum:2004fk}.
For that, in Ref.~\cite{Haidenbauer:2023qhf}
a novel regularization scheme, the so-called semilocal momentum space 
regularization, has been employed which has been 
already successfully applied in studies of the $NN$ interaction within chiral
effective field theory up to high orders \cite{Reinert:2017usi}.
The name refers to the fact that a local regulator is applied to the 
meson-exchange contributions whereas the contact terms, being 
nonlocal by themselves, are regularized with a nonlocal function.
The cutoff masses entering the regulator function 
considered in Ref.~\cite{Haidenbauer:2023qhf} are in
the range of $500$--$600$~MeV. For comparison, the cutoffs employed
in the SMS $NN$ interaction are in the range
$350$--$550$~MeV~\cite{Reinert:2017usi}.

With the NLO and N$^2$LO SMS $YN$ potentials 
an excellent description of the low-energy $\Lambda p$, $\Sigma^- p$, and
$\Sigma^+ p$ scattering cross sections could be achieved, resulting in a 
total $\chi^2$ of $15$--$16$ for the commonly considered $36$ data points \cite{Haidenbauer:2013oca}. Clearly, those data are primarily sensitive 
to the $YN$ interaction in the $S$ waves and, accordingly, only the 
LECs contributing to the $S$-wave potential could be reliably fixed. 
Nonetheless, in addition, 
new measurements of angular distributions for the $\Sigma N$ channels
from J-PARC \cite{J-PARCE40:2021qxa,J-PARCE40:2021bgw,J-PARCE40:2022nvq} 
have been analyzed in an attempt to determine also the strength of the 
contact interactions in the $P$ waves.

It is also noteworthy to mention that
the SMS $YN$ potentials have been utilized in calculations of light 
$\La$-hypernuclei, based on the Faddeev-Yakubovsky approach and the 
no-core shell model (so far up to $A\approx 8$). The
results, reported in Refs.~\cite{Le:2024rkd,Haidenbauer:2025zrr}, show 
that the separation energies based on the two-body interaction alone
are already fairly close to the experimental values
\cite{HypernuclearDataBase}.
When chiral ($\La NN$, $\Si NN$) 
three-body forces are included, which arise 
at N$^2$LO in the chiral expansion \cite{Petschauer:2015elq},
a quantitative agreement with the experimental separation
energies is achieved.  
 
Besides the SMS $YN$ potentials we will also present
results for the earlier NLO $YN$ potentials developed by 
the J\"ulich-Bonn-Munich and J\"ulich-Bonn groups,
NLO13 \cite{Haidenbauer:2013oca} and NLO19 \cite{Haidenbauer:2019boi},
that are based on a different regularization scheme. 
In those potentials a nonlocal regulator is used for all 
components, with a cutoff mass ranging from $500$ to $650$~MeV.  
We include here the results based on the NLO13 and NLO19
potentials with $500$~MeV. Those are the potentials employed
in the study of the hyperon puzzle by Gerstung \textit{et al.}
\cite{Gerstung:2020ktv}. Indeed, among the NLO13 and NLO19 
$YN$ potentials those with cutoff 500~MeV yield each the 
most attractive results for $U_\La$ at $\rho_0$. 
We note that the results of those potentials for the $\La p$ and $\Si N$
cross sections at low energies are practically identical to the
ones by the SMS $YN$ interaction~\cite{Haidenbauer:2023qhf,Haidenbauer:2013oca,Haidenbauer:2019boi}.

Regarding the $NN$ interaction we employ the corresponding 
SMS potentials by Reinert \textit{et al.}\cite{Reinert:2017usi} 
for consistency reasons. Specifically, we use the 
potentials with the highest order in the chiral expansion,
namely N$^4$LO$^+$ in the notation of Ref.~\cite{Reinert:2017usi}.

\subsection{Brueckner theory}

We utilize  conventional Brueckner theory to calculate
the single-particle potential of hyperons in nuclear matter
using the $YN$ interactions in free space.
Here, we briefly summarize the basics of the formalism and provide the 
essential formulas. A detailed description
can be found in Ref.~\cite{Petschauer:2015nea}; see also Refs.~\cite{Reuber:1993ip,Rijken:1998yy,Kohno:1999nz,Schulze:1998jf,Vidana:1999jm}.

Within the Brueckner theory, the two-body potential $V$ in free space is
converted into an effective two-body interaction in medium, the 
$G$ matrix, which is evaluated by solving the Bethe-Goldstone equation
\begin{align}
	\label{eq:Gmatrix}
	G(\omega) = V + V  \dfrac{Q}{\omega - H_0 + i\epsilon} G(\omega),
\end{align}
with the so-called starting energy $\omega$. The Pauli operator $Q$ eliminates
intermediate two-body states below the Fermi sea.
The energy denominator in Eq.~(\ref{eq:Gmatrix}) is the difference between
the starting energy $\omega$ and the total energy of
the intermediate state, which includes the baryon 
single-particle potentials 
\begin{eqnarray} \label{eq:omos}
 \omega &= E_{B_1}(k_{1})+E_{B_2}(k_{2})\,, \nonumber \\
 E_{B_i}(k_i) &= m_{B_i} + \displaystyle\frac{k_i^2}{2m_{B_i}} + \mathrm{Re} U_{B_i}(k_i) \,.
\end{eqnarray}
Here, the starting energy $\omega$ is usually chosen ``on-shell'',
i.e., it is set equal to the energy of the two particles in the
initial state. 
The angle average of the Pauli operator and the energy denominator
is adopted, so that after a standard partial-wave decomposition 
there remains only a one-dimensional integral equation.
Nonetheless, for the
hyperons $\La$ and $\Si$ the channel coupling 
$\La N$-$\Si N$ has to be taken into account. Thus, depending
on whether the $G$ matrix is calculated in the isospin or
particle basis, coupled-channel equations analogous to
Eq.~(\ref{eq:Gmatrix}) for two or three
channels have to be solved.
The single-particle potential of a baryon $B_1=(p,n,\La,\Si^{0,\pm})$
is typically chosen as the Hartree-Fock type one:
\begin{align}
	\label{eq:U_N}
	U_N(k) &= \sum_{m\, \in\, \text{Fermi sea}}
    \langle Nm| G |Nm \rangle -
        \langle Nm| G |mN \rangle, \\
    \label{eq:U_Y}
	U_Y(k) &= \sum_{m\, \in\, \text{Fermi sea}}
     \langle Ym| G |Ym \rangle.
\end{align}
Note that the Bethe-Goldstone equation~\eqref{eq:Gmatrix} and
the single-particle potentials $U_{B_i}$ must be evaluated self-consistently
because the single-particle potential appears in
Eq.~(\ref{eq:Gmatrix}) through the energy denominator.

The definition of the $U_{B_i}$ in Eqs.~\eqref{eq:U_N} and \eqref{eq:U_Y}
applies only to occupied states within the Fermi sea. 
The single-particle potential above the Fermi sea
can be chosen arbitrarily and should be chosen to achieve
faster convergence in the hole-line expansion.
There are commonly two choices of $U_{B_i}$: the gap choice which omits
$U$ above the Fermi sea and the continuous choice which uses $U$ as 
defined in Eqs.~\eqref{eq:U_N} and \eqref{eq:U_Y} also above the Fermi sea.
There are calculations that include higher-order contributions 
in the hole-line expansion~\cite{Song:1998zz,Baldo:2001mv,Lu:2017nbi}
and those observe a 
faster convergence by the continuous choice over the gap choice.
The exploratory calculation of the in-medium properties of the SMS $YN$ 
potentials in Ref.~\cite{Haidenbauer:2023qhf} was based on the gap choice.

In the actual calculations partial waves up to 
a total angular momentum of $J=5$ are taken into account.
Note that due to the local regulator used for the meson-exchange
contribution the potential drops off more slowly for large momenta,
which requires appropriate mesh points with
large maximum momentum
for calculating the $G$ matrix

In Ref.~\cite{Kohno:2018gby} some sort of oscillatory behavior of $U_N$
was found in calculations with the 
N$^3$LO $NN$ potential by Epelbaum et al. from 2005~\cite{Epelbaum:2004fk}
and a density-dependent $NN$ force~\cite{Kohno:2013ihv}. 
A similar behavior 
was also observed by Gerstung \textit{et al.}\cite{Gerstung:2020fzk}
for the N$^3$LO $NN$ potential by Entem and Machleidt~\cite{Entem:2003ft}.
In both calculations an additional cutoff regularization has been 
introduced and applied to the single-particle potential in the $G$-matrix 
equation to obtain numerically stable results.  
We do not observe such artificial oscillations for the 
SMS $NN$ potentials up to $\rho = 2\rho_0$
and, therefore, no additional cutoff is required.
In fact, also for the semilocal coordinate-space regularized (SCS)
$NN$ potentials employed 
in the studies of Hu \textit{et al.}\cite{Hu:2016nkw,Hu:2019zwa} no cutoff 
artifacts were found~\cite{Hu:private}. 

\section{Results}
\label{sec:results}

In this section we show and discuss results for the hyperon
single-particle potentials.
In particular, we document the momentum and density dependence of the $\Lambda$ and $\Sigma$ single-particle
potentials predicted by the SMS $YN$ interactions.
We start with some results for nuclear matter
based on the SMS $NN$ potentials~\cite{Reinert:2017usi}.
Specifically we show the predicted nucleon single-particle
potential $U_N$.
As one can see from Eq.~(\ref{eq:omos}), this quantity
enters as input in the calculation of the in-medium 
properties of the hyperons. 

\subsection{Nuclear matter}
\label{sec:nuclMat}

Since to the best of our knowledge nuclear matter results for the 
SMS $NN$ potentials of Reinert \textit{et al.}\cite{Reinert:2017usi} have so
far not been reported, we present such results in this subsection.
We focus on the potentials of highest order in the chiral expansion,
namely N$^4$LO$^+$, which we use for our calculation of the
hyperon properties. 
For comparison we present also results for the N$^3$LO potentials by Entem 
and Machleidt (EM) \cite{Entem:2003ft}, which have been used by  Gerstung
\textit{et al.}~\cite{Gerstung:2020ktv}. It should be said that the actual calculations
in that reference include also an additional effective density-dependent $NN$ 
interaction, derived from N$^2$LO $NNN$ forces following the prescription 
of Holt \textit{et al.}\cite{Holt:2009uk,Holt:2009ty}.

\begin{figure}[tbp]
    \centering
    \includegraphics[width=0.85\linewidth]{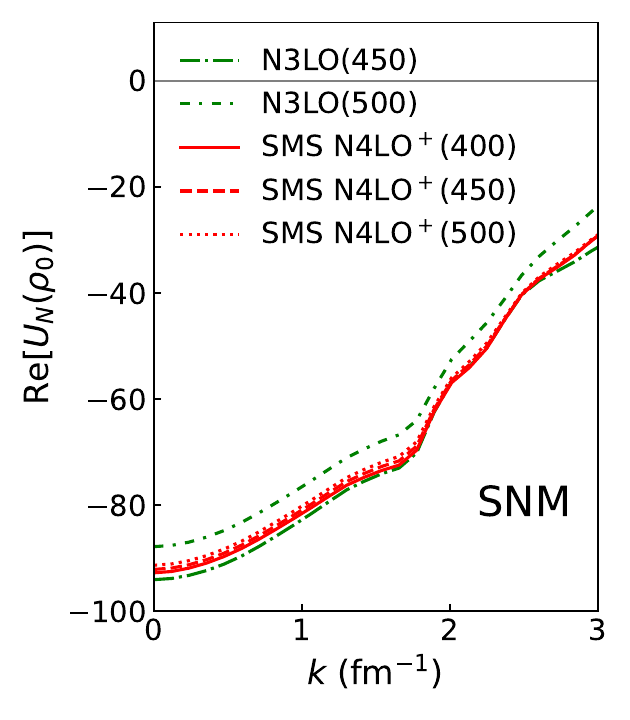}
    \includegraphics[width=0.85\linewidth]{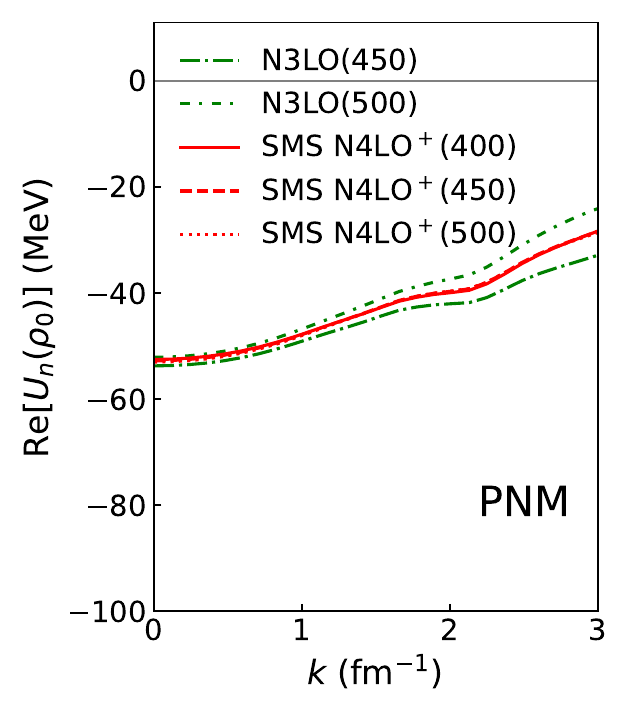}
    \caption{Nucleon single-particle potentials in symmetric nuclear matter (top) and pure neutron matter (bottom) at $\rho_0$.
    N$^3$LO(450) and N$^3$LO(500) refer to the N$^3$LO NN potential by Entem and Machleidt~\cite{Entem:2003ft} with corresponding cutoff,
    while the SMS N$^4$LO$^+$ potential is by Reinert 
    \textit{et al.}\cite{Reinert:2017usi}.
    }
    \label{fig:UN}
\end{figure}

Nucleon single-particle potentials in SNM and in PNM at nuclear 
matter saturation density
are shown in Fig.~\ref{fig:UN},  
for the EM potentials~\cite{Entem:2003ft}
with cutoffs 450 and 500 MeV 
and for the SMS N$^4$LO$^+$ potentials with cutoffs
$400$--$500$~MeV~\cite{Reinert:2017usi}.
Remarkably the momentum dependence of $U_N$ resulting from
the SMS $NN$ potentials exhibits basically no variation with 
the cutoff, testifying that the residual regulator
dependence at that order of the chiral expansion is 
indeed very small. In the case of the EM N$^3$LO potential
the cutoff variation leads to some spread in the 
value of $U_N$; however, also here the actual momentum 
dependence is very similar.

\begin{figure}[tbp]
    \centering
    \includegraphics[width=0.90\linewidth]{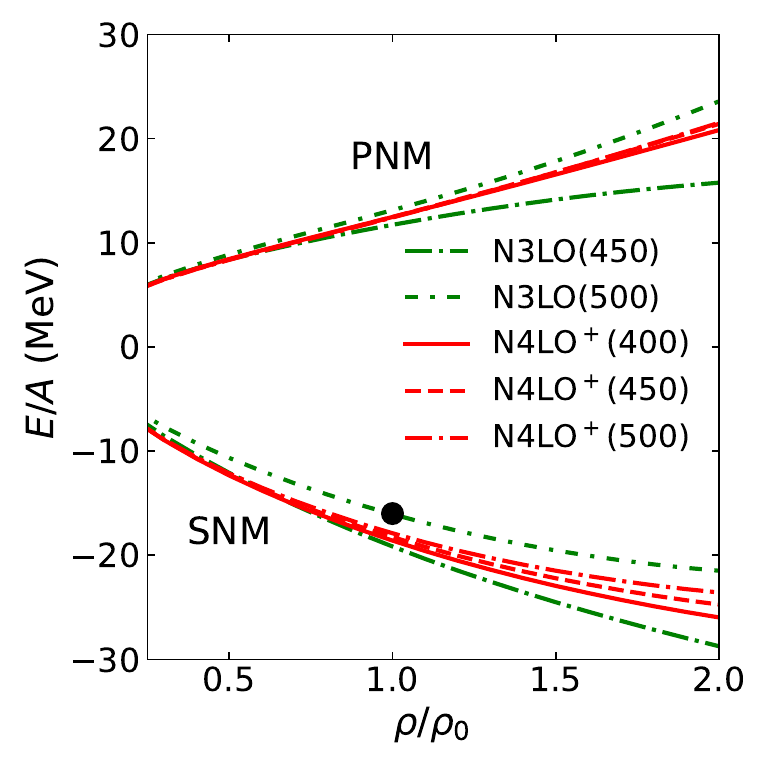}
    \caption{Total energy per particle in both symmetric nuclear matter
    and pure neutron matter. Results are shown for the N$^3$LO~\cite{Entem:2003ft} and
    SMS N$^4$LO$^+$~\cite{Reinert:2017usi} $NN$ potentials.
    The empirical value at the nuclear matter saturation point, $E/A=-16$~MeV, is indicated by a circle.
    }
    \label{fig:EA}
\end{figure}

In the BHF approximation the total energy is given by  
\begin{align}
    E = \sum_{m \in \text{Fermi sea}}
    \left[\langle m | T | m \rangle + \frac{1}{2}\,U_N(k_m)\right],
\end{align}
where $T$ is the kinetic energy operator.
In Fig.~\ref{fig:EA} results
for the total energy per particle
in both SNM and in PNM are shown up to $2\rho_0$,
corresponding to the Fermi momentum $k_F \le 1.70$~fm$^{-1}$ for SNM
and $k_F \le 2.15$~fm$^{-1}$ for PNM.
The Fermi momentum is related to the density via
$\rho = 2k^3_F/(3\pi^2)$ for SNM and $\rho = k^3_F/(3\pi^2)$ for PNM,
respectively.
Both the EM N$^3$LO and the SMS N$^4$LO$^+$ $NN$ potentials yield similar results,
expectedly because the underlying single-particle potentials are also very similar.
The SNM results for N$^4$LO$^+$ are slightly too attractive and lie below
the empirical value at the saturation point. 
In the case of the EM potentials there is a larger variation with the cutoff. 

\subsection{$\La$ in symmetric nuclear matter}
\label{sec:La}

\begin{table*}[tbp]
\renewcommand{\arraystretch}{1.2}
\centering
\caption{
Partial-wave contributions to $ U_\La (k = 0)$ (in MeV)
{at} $k_F = 1.35 \ {\rm fm}^{-1}$. Results are presented for 
the SMS $YN$ interactions at different chiral orders and with 
different cutoffs \cite{Haidenbauer:2023qhf}
and, in addition, for the NLO13(500) \cite{Haidenbauer:2013oca}
and NLO19(500) \cite{Haidenbauer:2019boi}  interactions. 
For reference, selected results based on the gap choice are included too. 
}
\begin{tabular}{c|rcrrrc|c}
\hline\hline
 & $^1S_0$ & $^3S_1+^3D_1$ & $^3P_0$ & $^1P_1$ & $^3P_1$ & $^3P_2+^3F_2$ &
\, Total \, \\
\hline
SMS LO(700)      & $-13.9$ & $-28.2$ & $-1.0$ & $0.7$ & $1.0$ & $-0.8$ & $-42.4$ \\ 
\hline
SMS NLO(500)     & $-15.7$ & $-25.6$ & $0.4$ & $4.5$ & $2.0$ & $-3.2$ & $-37.7$ \\ 
SMS NLO(550)     & $-15.0$ & $-24.5$ & $0.5$ & $2.5$ & $1.5$ & $-4.1$ & $-39.3$ \\ 
SMS NLO(600)     & $-12.5$ & $-20.2$ & $0.4$ & $1.0$ & $1.5$ & $-4.0$ & $-34.0$ \\ 
\hline
SMS N$^2$LO(500)    & $-16.2$ & $-24.8$ & $0.3$ & $2.9$ & $0.6$ & $-3.2$ & $-41.2$ \\ 
SMS N$^2$LO(550)$^a$  & $-15.2$ & $-26.9$ & $0.1$ & $1.2$ & $0.9$ & $-4.9$ & $-45.8$ \\ 
SMS N$^2$LO(550)$^b$  & $-15.1$ & $-27.3$ & $0.4$ & $1.5$ & $1.6$ & $-3.4$ & $-43.4$ \\ 
SMS N$^2$LO(600)    & $-15.3$ & $-25.4$ & $0.2$ & $1.6$ & $1.0$ & $-5.0$ & $-44.1$ \\ 
\hline
NLO13(500)       & $-15.3$ & $-18.9$ & $0.9$ & $0.2$ & $1.6$ & $-1.3$ & $-33.1$ \\ 
NLO19(500)       & $-13.5$ & $-28.7$ & $0.9$ & $0.3$ & $1.6$ & $-1.2$ & $-40.9$ \\ 
\hline
\hline
SMS LO(700)     gap & $-12.1$ & $-24.8$ & $-0.9$ & $0.7$ & $1.0$ & $-0.7$ & $-37.0$ \\ 
SMS NLO(500)    gap & $-15.2$ & $-22.2$ & $0.5$ & $5.0$ & $2.1$ & $-3.0$ & $-32.9$ \\ 
SMS N$^2$LO(500)   gap & $-15.6$ & $-20.8$ & $0.4$ & $3.3$ & $0.6$ & $-3.0$ & $-35.9$ \\ 
NLO13(500)      gap & $-14.7$ & $-13.7$ & $1.0$ & $0.3$ & $1.6$ & $-1.2$ & $-27.0$ \\ 
NLO19(500)      gap & $-12.0$ & $-27.1$ & $1.0$ & $0.3$ & $1.7$ & $-1.2$ & $-37.5$ \\ 
\hline\hline
\end{tabular}
\label{tab:La}
\renewcommand{\arraystretch}{1.0}
\end{table*}

We now proceed to the in-medium properties of the hyperons.
The inputted $U_N$ is calculated using the SMS N$^4$LO$^+$ $NN$
force with the $450$~MeV cutoff.
First, we focus on the $\La$ single-particle potential
at nuclear matter saturation density and for zero momentum,
which can be compared with quasiempirical values
inferred from hypernuclear experiments. 
Table~\ref{tab:La} summarizes the results for $U_\Lambda(k=0)$ 
and its partial-wave decomposition. We adopt the notation from
Ref.~\cite{Haidenbauer:2023qhf} and specify the various potentials 
by their chiral order and by the employed cutoff mass. 
The values of $U_\Lambda(k=0)$ predicted by the 
SMS N$^2$LO potentials range
from $-46$ to $-41$~MeV; those of the NLO potentials are between
$-39$ and $-34$~MeV.
In comparison, the result for NLO19 ($-41$~MeV) is right in the middle
of the SMS results, while the one for NLO13 ($-33$~MeV) is at the 
upper edge. 
Also phenomenological $YN$ models predict results for $U_\Lambda$ in that
range~\cite{Rijken:1998yy,Reuber:1993ip,Nagels:2015lfa}.
The usually cited quasiempirical value, i.e., the well depths of a 
$\La$-nucleus Woods-Saxon potential fixed in an analysis of 
hypernuclear data, amounts to $U_\Lambda\approx-30$~MeV \cite{Gal:2016boi}.
A recent phenomenological analysis~\cite{Friedman:2023ucs} based
on hypernuclei estimates that the $YN$ two-body contribution to
$U_\Lambda (k=0)$ is $-38.6 \pm 0.8$~MeV and attributes a contribution
of $11.3\pm 1.4$~MeV to repulsive three-body forces. 

It is interesting to see that the theoretical predictions from chiral $YN$ 
interactions and the two-body result from the phenomenological analysis 
match so well.
Nonetheless, one should keep in mind that contributions from three-body 
forces are in general scheme dependent. However, within the Weinberg scheme
they can be made consistent with the two-body forces using gradient flow regularization~\cite{Krebs:2023gge}. Such work is underway by the LENPIC
Collaboration. 

For testing purposes, we performed selected calculations
with the gap choice. With that choice for the energy in the intermediate 
state, the single-particle potential $U_\La$ at $\rho_0$ 
is $3$ to $7$~MeV less attractive, well in line with earlier observations
\cite{Rijken:1998yy}. 
Also, our results with the gap choice agree within $1.2$~MeV with the 
ones reported in Ref.~\cite{Haidenbauer:2023qhf},
which utilized a parametrization of $U_N$ results for the Argonne AV18 $NN$
potential~\cite{Wiringa:1994wb}, provided in Ref.~\cite{Isaule:2016pnn}.
On the other hand, our result for NLO13(500) with continuous choice 
is clearly different from that in Ref.~\cite{Petschauer:2015nea}. 
However, in that work only an NLO $NN$ potential has been used.
Potentials up to that order of the chiral expansion tend to be too 
repulsive for higher momenta/energies~\cite{Petschauer:2015nea}.   
Still, the comparison illustrates that the nucleon single-particle 
potential can affect $U_Y(k=0)$.
That said, when using the N$^3$LO $NN$ potentials from EM or the AV18 $NN$,
we obtain results for $U_Y$ very similar to the calculation 
with the SMS N$^4$LO$^+$ $NN$ interaction.

We want to remind the reader that at LO only a very basic description of 
the $YN$ interaction can be obtained~\cite{Haidenbauer:2023qhf}. We
include the corresponding results primarily for completeness and
because they enter into the uncertainty estimate. 

\begin{table*}[tbhp]
\renewcommand{\arraystretch}{1.2}
\centering
\caption{
Partial-wave contributions to $U_\Lambda(k=0)$ (in MeV) in symmetric nuclear matter
for $\rho/\rho_0=(0.5,1.0,2.0)$, corresponding to Fermi momenta
$k_F=(1.07,1.35,1.7)~{\rm fm}^{-1}$.
}
\begin{tabular}{c|c|rrrrrr|r|r|r|r}
\hline\hline
& \ $\rho/\rho_0$ \ & $^1S_0$ & $^3S_1+^3D_1$ & $^3P_0$ & $^1P_1$ & $^3P_1$ & $^3P_2+^3F_2$ & \ $S$ waves \ & \ $P$ waves \ & \ $D$ waves \ & \, Total \, \\
\hline
SMS LO(700)     & 0.5 & $-$8.1 & $-$16.9 & $-$0.4 & 0.2 & 0.3 & $-$0.3 & $-$25.0 & $-$0.2 & $-$0.1 & $-$25.2 \\
     & 1.0 & $-$13.9 & $-$28.2 & $-$1.0 & 0.7 & 1.0 & $-$0.8 & $-$42.1 & $-$0.2 & $-$0.2 & $-$42.4 \\
     & 2.0 & $-$23.3 & $-$44.6 & $-$2.4 & 1.8 & 2.9 & $-$1.9 & $-$68.2 & 0.3 & $-$0.9 & $-$68.6 \\
\hline
SMS NLO(550)    & 0.5 & $-$9.1 & $-$17.0 & 0.1 & 0.7 & 0.5 & $-$1.4 & $-$26.1 & $-$0.1 & $-$0.1 & $-$26.2 \\
    & 1.0 & $-$15.0 & $-$24.5 & 0.5 & 2.5 & 1.5 & $-$4.1 & $-$39.6 & 0.4 & $-$0.2 & $-$39.3 \\
    & 2.0 & $-$24.1 & $-$26.6 & 1.8 & 8.4 & 4.7 & $-$11.3 & $-$50.8 & 3.6 & $-$0.8 & $-$48.0 \\
\hline
SMS N$^2$LO(550)$^a$ & 0.5 & $-$9.2 & $-$18.2 & $-$0.0 & 0.3 & 0.2 & $-$1.8 & $-$27.4 & $-$1.3 & $-$0.3 & $-$28.9 \\
 & 1.0 & $-$15.2 & $-$26.9 & 0.1 & 1.2 & 0.9 & $-$4.9 & $-$42.0 & $-$2.6 & $-$1.1 & $-$45.8 \\
 & 2.0 & $-$24.2 & $-$31.8 & 0.6 & 4.9 & 3.4 & $-$12.3 & $-$55.5 & $-$3.3 & $-$4.4 & $-$63.9 \\
\hline
SMS N$^2$LO(550)$^b$ & 0.5 & $-$9.2 & $-$18.3 & 0.1 & 0.4 & 0.5 & $-$1.3 & $-$27.4 & $-$0.4 & $-$0.3 & $-$28.0 \\
 & 1.0 & $-$15.1 & $-$27.3 & 0.4 & 1.5 & 1.6 & $-$3.4 & $-$42.4 & 0.1 & $-$1.1 & $-$43.4 \\
 & 2.0 & $-$24.0 & $-$33.0 & 1.6 & 5.7 & 5.7 & $-$8.5 & $-$56.7 & 4.7 & $-$4.4 & $-$56.9 \\
\hline\hline
NLO13(500)      & 0.5 & $-$9.3 & $-$15.3 & 0.2 & 0.1 & 0.5 & $-$0.5 & $-$24.6 & 0.3 & $-$0.1 & $-$24.3 \\
      & 1.0 & $-$15.3 & $-$18.9 & 0.9 & 0.2 & 1.6 & $-$1.3 & $-$34.3 & 1.5 & $-$0.2 & $-$33.1 \\
      & 2.0 & $-$22.8 & 0.3 & 3.5 & 0.8 & 4.9 & $-$3.5 & $-$22.9 & 5.7 & $-$0.8 & $-$17.9 \\
\hline
NLO19(500)      & 0.5 & $-$8.7 & $-$18.3 & 0.2 & 0.1 & 0.5 & $-$0.5 & $-$27.0 & 0.4 & $-$0.1 & $-$26.6 \\
      & 1.0 & $-$13.5 & $-$28.7 & 0.9 & 0.3 & 1.6 & $-$1.2 & $-$42.3 & 1.5 & $-$0.2 & $-$40.9 \\
      & 2.0 & $-$15.6 & $-$36.5 & 3.6 & 0.8 & 5.0 & $-$3.3 & $-$52.5 & 6.1 & $-$0.8 & $-$47.2 \\
\hline\hline
\end{tabular}
\label{tab:La_rho}
\renewcommand{\arraystretch}{1.0}
\end{table*}

Let us now discuss the results in Table~\ref{tab:La} in detail. 
Obviously, there is a sizable variation in the contribution from the
$^3S_1-^3D_1$ partial wave, within the SMS potentials but also in comparison
to NLO13 and NLO19. It is in the order of 25~\%. 
The $^1S_0$ contributions vary in the order of 10~\%. 
There are also noticeable variations in the $P$-wave contributions.
As explained in the Introduction, the $P$ waves have been determined
differently in the SMS $YN$ potentials and in NLO13 and NLO19. 
This is clearly reflected in the results in Table~\ref{tab:La}. While there is
some variation in the $P$-wave contributions for the SMS potentials, there are 
more drastic differences to those for NLO13
and NLO19. As one can see, they differ by a factor of $2$--$3$ ($^3P_0$, $^3P_2$) 
or even by a factor of $5$--$10$ or more ($^1P_1$). 
In this context, let us mention that calculations with a $YN$ potential 
derived within covariant chiral EFT~\cite{Zheng:2025sol} predict even 
larger (and repulsive) $P$-wave contributions at $\rho_0$. Note, however, 
that the study is performed within the Dirac Brueckner-Hartree-Fock approach.
Finally, we want to remark that the $P$-wave interactions of the NLO13
and NLO19 potentials are identical. In contrast, in the case of 
SMS N$^2$LO(550)$^a$ and N$^2$LO(550)$^b$ the $P$ waves are different while 
the $S$-wave interactions are the same. 
The two variants are the result of different fits to data on the 
$\Si^+p$ differential cross section \cite{Haidenbauer:2023qhf}
and reflect remaining ambiguities in the $P$-wave contributions.

In Table~\ref{tab:La_rho}, we list again the partial-wave contributions, 
however, from a different perspective, namely as a function of the density.
We restrict ourselves to the SMS potentials with $550$~MeV cutoff and 
NLO13 and NLO19. 
One can see that the total $S$-wave contribution clearly dominates $U_\La$
up to $2\rho_0$. The difference between NLO and N$^2$LO is fairly
moderate. The SMS results for the $S$ waves are also similar to that of 
NLO19.

With regard to the $P$-wave contributions, those are overall small but
they become more relevant with increasing density. Their smallness is 
partly due to significant cancellations between the individual contributions.
Comparing the individual $P$-wave results for SMS N$^2$LO and NLO  
with those from NLO13 and NLO19 reveals that there is no clear tendency.
This reflects the fact that the $\La N$ $P$-wave interactions are 
only poorly constrained by the existing experimental data. 
Specifically instructive are the results for the potentials N$^2$LO(550)$^a$
and N$^2$LO(550)$^b$ established by alternative fits 
to the $\Si^-p$ differential cross section, as already mentioned above. 
In this case, even the signs of the total $P$-wave contribution differ 
at $\rho_0$ and $2\rho_0$.
A discussion of the impact of the $P$-wave ambiguities on $U_\La$
in the context of the $YN$ potentials of the Nijmegen Group can be 
found in Ref.~\cite{Rijken:1998yy}.

\begin{figure*}[htbp]
    \centering
    \includegraphics[width=0.99\linewidth]{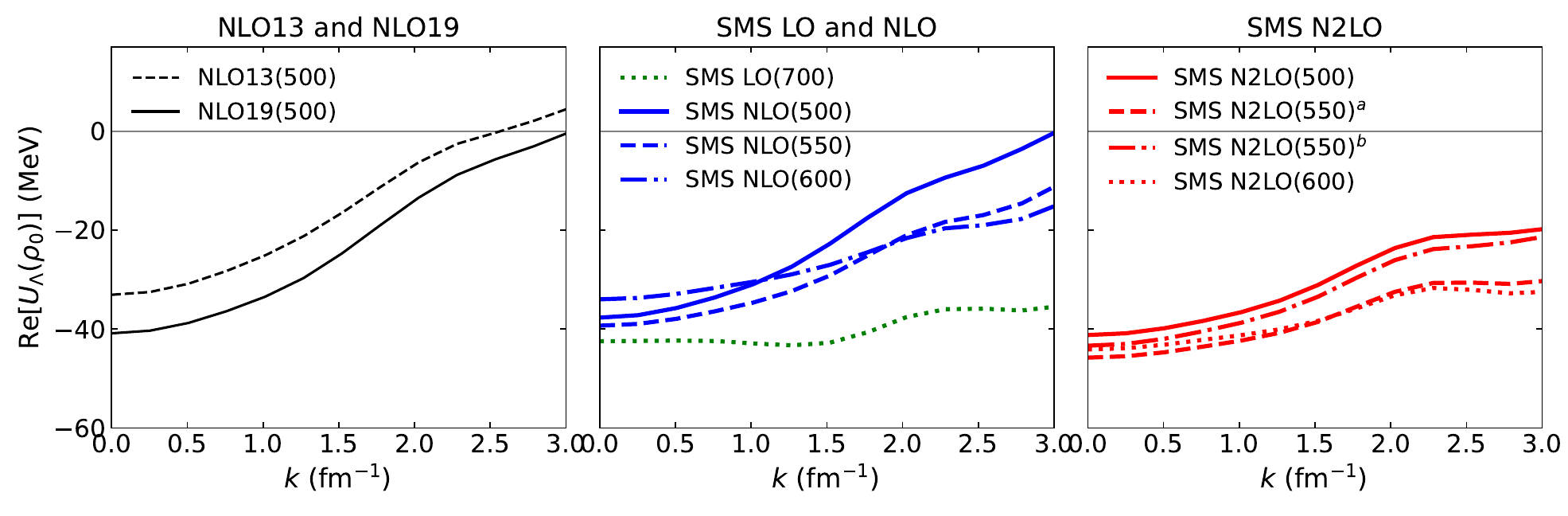}
    \caption{Momentum dependence of the $\La$ single-particle potentials
    in symmetric nuclear matter at $\rho_0$.
    }
    \label{fig:ULmom_SNM}
\end{figure*}

Table~\ref{tab:La_rho} reveals that there are also sizable contributions 
from the $D$ waves, specifically for the N$^2$LO potentials at $2\rho_0$.
In order to understand that one has to recall that the contributions 
at N$^2$LO which arise from two-meson exchange with subleading 
meson-baryon vertices are relatively large \cite{Epelbaum:2004fk}.  
However, unlike in $S$ and $P$ waves, there are no contact terms in 
the $D$ waves yet which would counterbalance their effect and/or
allow one to reduce possible artifacts from the employed regulator.  
Such contact terms arise first at N$^3$LO in the chiral expansion. 
In fact, in this context it is very instructive to look at the 
corresponding $NN$ results by Hu \textit{et al.}\cite{Hu:2016nkw}; see their 
Fig.~1. In the nuclear matter case, the N$^2$LO result clearly deviates 
from the others and becomes exceptionally attractive with increasing density. 
When going to N$^3$LO the general trend is restored. It is likely 
that the same will happen also in the case of the in-medium properties
of hyperons. Unfortunately, due to the lack of appropriate scattering
data there is no way to extend the $YN$ interaction to N$^3$LO in
the foreseeable future.

In Fig.~\ref{fig:ULmom_SNM} predictions for the momentum dependence of 
$U_\La$ at $\rho_0$ are shown. 
The momentum dependence of $U_\Lambda$ from microscopic calculations
is often used as a guideline for establishing mean-field models. 
For example, in Ref.~\cite{Chorozidou:2024gyy}, it is argued
that the momentum dependence can be a key to solve the hyperon puzzle 
of neutron stars.
Furthermore, the momentum dependence has an impact on
the hyperon dynamics in heavy-ion collisions~\cite{Nara:2022kbb}.

In the case of the NLO13 and NLO19 potentials, the resulting $U_\La(k)$ rise  
monotonically with increasing momentum, with a very similar trend.
The SMS potentials exhibit a noticeably weaker momentum dependence,
except for SMS NLO(500), where the behavior is comparable to that for  
NLO19(500).
Overall, the momentum dependence predicted for the N$^2$LO potentials is
more moderate than those for the NLO interactions. 
The variation of $U_\La(k)$ with the cutoff is of similar magnitude for
the SMS NLO and N$^2$LO potentials. Also, in both cases there is a 
slight increase with increasing momentum. 

It is instructive to compare the variation of the momentum dependence
with the cutoff to the one observed for $U_N$ based on the N$^4$LO$^+$
potentials; see Fig.~\ref{fig:UN}. It shows what one can expect in a case 
where the chiral expansion is already well converged and the residual 
regulator dependence is minimal. 

\begin{figure}
    \centering
    \includegraphics[width=0.99\linewidth]{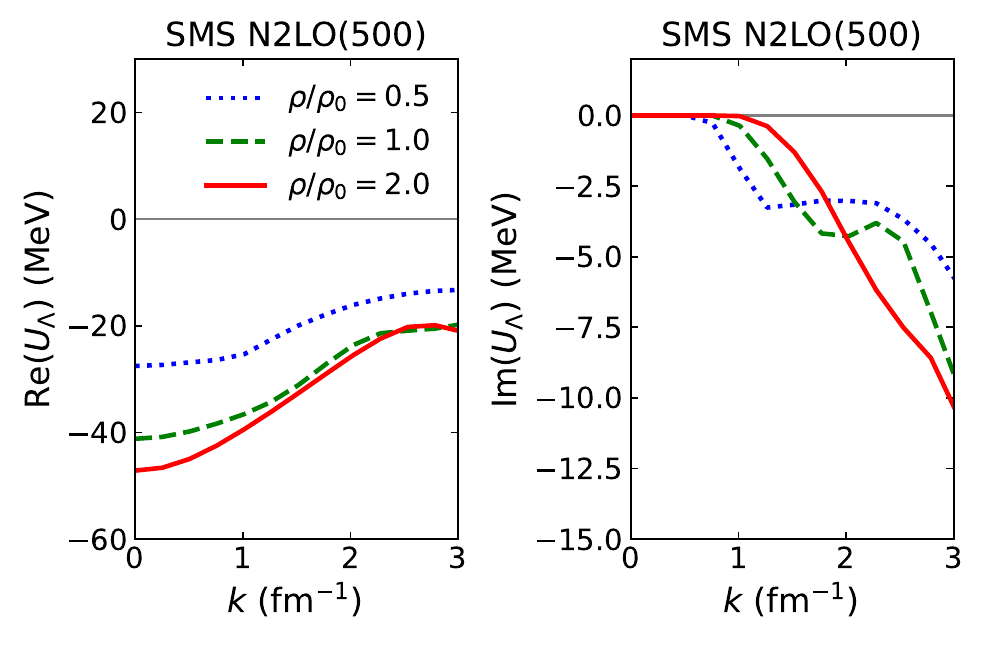}
    \caption{Momentum dependence of $U_\La$ in symmetric nuclear matter
    for $\rho/\rho_0=(0.5,1.0,2.0)$, corresponding
    to the Fermi momenta $k_F=(1.07,1.35,1.7)~{\rm fm}^{-1}$.
    The SMS N$^2$LO(500) $YN$ potential is employed.
    }
    \label{fig:UYmom_rho}
\end{figure}

An important quantity related to the momentum dependence
is the effective mass, which is sensitive to the energy
difference between orbitals of the $\La$ as considered from mean-field models~\cite{Yamamoto:1988qz}.
The definition of the effective mass of $\La$ is usually given as
\begin{align}
    \dfrac{m^{*}_\Lambda}{m_\Lambda} = 
    \left[1 + 2m_\Lambda
    \dfrac{\partial \, \mathrm{Re} U_\La(k)}{\partial k^2}\Big|_{k=0}\right]^{-1}.
\end{align}
Since for the considered $YN$ potentials all $U_\Lambda$ are in line with 
a $k^2$ dependence below $k=1$~fm$^{-1}$, we evaluate the effective mass as~\cite{Kohno:1999nz}
\begin{align}
    \dfrac{m^{*}_\Lambda}{m_\Lambda} = 
    \left[1 + \dfrac{2m_\Lambda}{k^2}
    \left[U_\Lambda(k) - U_\Lambda(0)\right]
    \right]^{-1},
\end{align}
with $k=1$~fm$^{-1}$. The calculated values are listed in 
Table~\ref{tab:Meff}.
The values for the SMS N$^2$LO (NLO) potentials are around $0.79$--$0.86$
($0.73$--$0.83$), while those from NLO13 and NLO19 are around $0.7$.
For the SMS LO potential, $U_\Lambda(k)$ is practically constant for 
small $k$ and, accordingly, the effective mass is close to $1.0$.
The SMS NLO and N$^2$LO values are within the hypernuclear
constraint, $0.65 < m^*_\Lambda / m_\Lambda < 0.95$~\cite{Jinno:2023xjr}.

\begin{figure*}[tbhp]
    \centering
    \includegraphics[width=0.99\linewidth]{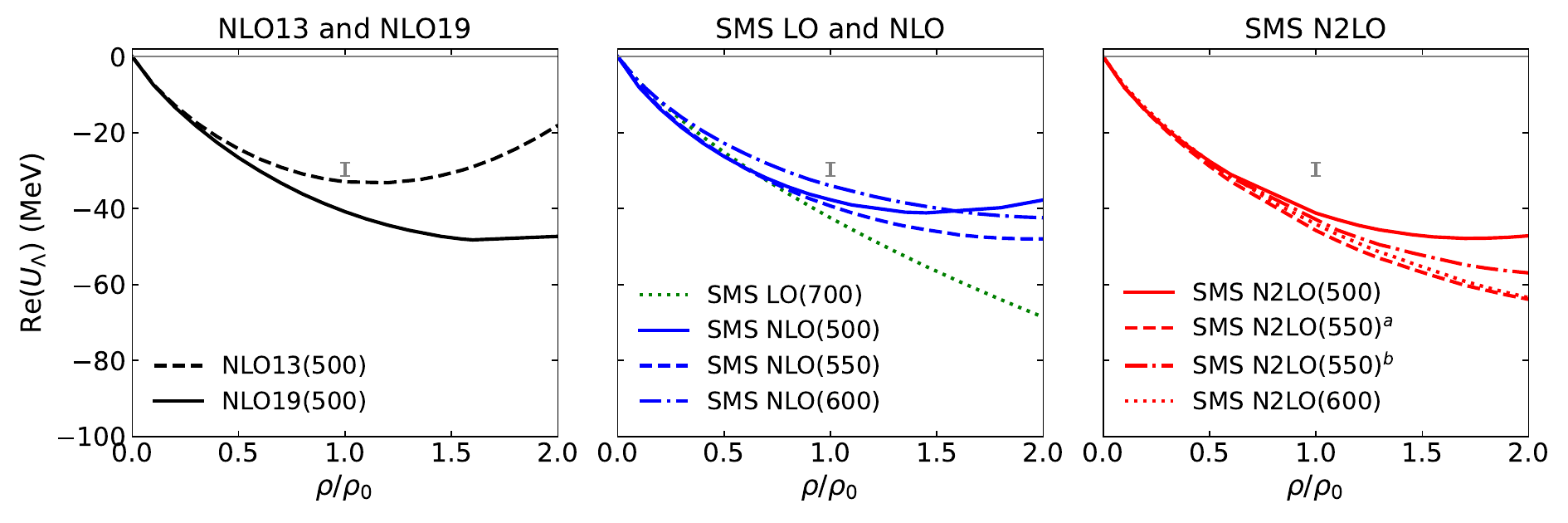}
    \caption{Density dependence of the $\Lambda$ single-particle potential in
    symmetric nuclear matter.
    The bar symbolizes the quasiempirical value~\cite{Gal:2016boi}. 
    }
    \label{fig:UL_rhoSNM}
\end{figure*}

\begin{table}[tbp]
\centering
\caption{
Normalized effective mass $m^*_\Lambda/m_\Lambda$ in symmetric nuclear matter at $\rho_0$.
}
\begin{tabular}{c|c}
\hline\hline
 & $m^*_\La/m_\La $ \\
\hline
SMS LO(700)      & $1.03$ \\
SMS NLO(500)     & $0.73$ \\
SMS NLO(550)     & $0.79$ \\
SMS NLO(600)     & $0.83$ \\
SMS N$^2$LO(500)    & $0.79$ \\
SMS N$^2$LO(550)$^a$ & $0.84$ \\
SMS N$^2$LO(550)$^b$  & $0.79$ \\
SMS N$^2$LO(600)    & $0.86$ \\
\hline
NLO13(500)       & $0.69$ \\
NLO19(500)       & $0.71$ \\
\hline\hline
\end{tabular}
\label{tab:Meff}
\end{table}

The variation in the momentum dependence with density is
presented in Fig.~\ref{fig:UYmom_rho}, exemplary for the SMS 
N$^2$LO(500) potential. 
We consider densities $\rho = 0.5\rho_0,\, \rho_0,\, 2\rho_0$, corresponding
to the Fermi momenta $k_F = (1.07, 1.35, 1.7)~{\rm fm}^{-1}$.
For the $\Lambda$ single-particle potential, the curvature around
$k=0$~fm$^{-1}$ remains essentially unchanged with increasing
density. This behavior means that the $\Lambda$ effective mass
decreases as the density increases, similar to the NLO13
results~\cite{Petschauer:2015nea}. 
The imaginary part of $U_\Lambda$ is also  shown in Fig.~\ref{fig:UYmom_rho}.
As one can see the actual momentum dependence changes noticeably with increasing 
density. 

Finally, in Fig.~\ref{fig:UL_rhoSNM} we show the density
dependence of $U_\La$ for SNM.
For the SMS interactions with a cutoff of $500$~MeV, and also for NLO13 and 
NLO19, $U_\La$ reaches a minimum in the considered density range. 
In the case of NLO13 there is even a pronounced rise from around $1.5\rho_0$
onwards, while a slower variation with density is seen for the 
other interactions. As known from the work of Gerstung \textit{et al.},
$U_\La$ for NLO13
eventually changes sign and becomes repulsive around $2.5\rho_0$~\cite{Gerstung:2020ktv}. 
The origin of the strikingly different behavior of the $U_\La$ results 
for NLO13 and NLO19 has been thoroughly discussed in
Ref.~\cite{Haidenbauer:2019boi}. 
It is a consequence of noticeable differences in the strength of the 
$\La N$-$\Si N$ transition potential between the two sets of $YN$ interactions.
Broadly speaking, a larger (smaller) channel-coupling strength leads to 
a less attractive (more attractive) result for $U_\La$, even when the 
corresponding $YN$ scattering results, including the $\La N$-$\Si N$
transition cross sections, are identical. 
The sensitivity of $U_\La$ to the $\La N$-$\Si N$ 
transition strength has been known for a long time
\cite{Nogami:1970rk,Bodmer:1971eh,Dabrowski:1973enk}.
For a complementary demonstration of coupled-channel effects
on the in-medium properties of the $\La$ in the context of the NLO13 $YN$
potential, we refer the reader to the work of Kohno~\cite{Kohno:2018gby}.
As likewise discussed in Ref.~\cite{Haidenbauer:2019boi}, the actual 
strength of the transition potential is closely interrelated 
with possible contributions from three-body forces.
Anyway, 
regarding the SMS NLO and N$^2$LO interactions with larger cutoffs, their
trends are still similar to that of the potentials with $500$~MeV but, of course,
the minimum of $U_\La$ will be reached at somewhat higher density.

\begin{table*}[tbhp]
\renewcommand{\arraystretch}{1.2}
\centering
\caption{Partial-wave contributions to the real part of $U_\Sigma(k=0)$ 
(in MeV) at $\rho_0$. 
}
\begin{tabular}{c|rrrrrr|r}
\hline\hline
& $^1S_0$ & $^3S_1+^3D_1$ & $^3P_0$ & $^1P_1$ & $^3P_1$ & $^3P_2+^3F_2$ &
\, Total \, \\
\hline
SMS LO(700)      & $-3.3$ & $8.0$ & $-1.0$ & $-0.3$ & $-0.4$ & $-1.8$ & $1.0$ \\
SMS NLO(500)     & $-3.4$ & $-7.4$ & $0.8$ & $0.4$ & $1.7$ & $-1.5$ & $-9.8$ \\
SMS NLO(550)     & $-2.7$ & $-4.6$ & $0.8$ & $-0.5$ & $0.6$ & $-4.5$ & $-11.3$ \\
SMS NLO(600)     & $-2.4$ & $-3.1$ & $0.7$ & $-1.0$ & $0.5$ & $-5.2$ & $-10.9$ \\
SMS N$^2$LO(500)    & $-5.9$ & $-3.2$ & $1.0$ & $0.6$ & $0.8$ & $0.3$ & $-7.8$ \\
SMS N$^2$LO(550)$^a$  & $-3.4$ & $-4.9$ & $0.5$ & $-0.8$ & $1.0$ & $-1.6$ & $-10.3$ \\
SMS N$^2$LO(550)$^b$  & $-3.5$ & $-4.9$ & $0.8$ & $-3.3$ & $1.5$ & $-0.4$ & $-11.0$ \\
SMS N$^2$LO(600)    & $-3.2$ & $-5.3$ & $0.6$ & $-2.7$ & $1.2$ & $-2.0$ & $-12.7$ \\
\hline
NLO13(500)       & $-4.9$ & $6.4$ & $1.3$ & $0.2$ & $1.2$ & $0.2$ & $3.7$ \\
NLO19(500)       & $-4.4$ & $13.1$ & $1.3$ & $0.2$ & $1.2$ & $0.2$ & $11.2$ \\
\hline \hline
SMS LO(700)     gap & $-2.3$ & $14.1$ & $-1.0$ & $-0.2$ & $-0.4$ & $-1.6$ & $8.3$ \\
SMS NLO(500)    gap & $-2.7$ & $-1.3$ & $0.8$ & $0.5$ & $1.7$ & $-1.2$ & $-2.4$ \\
SMS N$^2$LO(500)   gap & $-5.4$ & $6.2$ & $1.0$ & $0.6$ & $0.7$ & $0.6$ & $2.8$ \\
NLO13(500)      gap & $-4.3$ & $16.5$ & $1.4$ & $-0.0$ & $1.1$ & $0.2$ & $14.6$ \\
NLO19(500)      gap & $-3.8$ & $21.2$ & $1.3$ & $-0.1$ & $1.1$ & $0.1$ & $19.7$ \\
\hline\hline
\end{tabular}
\label{tab:Si}
\renewcommand{\arraystretch}{1.0}
\end{table*}

\begin{table*}[htbp]
\renewcommand{\arraystretch}{1.2}
\centering
\caption{
Partial-wave contributions to 
${\rm Re}\,U_\Sigma(k=0)$ (in MeV). Total results for the
isoscalar ($U^0_\Sigma$) and isovector ($U^1_\Sigma$) decomposition according
to Eq.~\eqref{eq:US_isospin} are given as well.
The column on the very right is the conversion width in nuclear matter, 
$\Gamma_\Si = -2\,{\rm Im}\,U_\Sigma(k=0)$.
Same description of interactions as in Table \ref{tab:La}.
}
\begin{tabular}{c|rrrrrr|rr|r}
\hline\hline
& \multicolumn{3}{c}{Isospin $I=1/2$} & \multicolumn{3}{c|}{Isospin $I=3/2$} & & & \\
& $^1S_0$ & $^3S_1$+$^3D_1$ & $P$ & $^1S_0$ & $^3S_1$+$^3D_1$ & $P$ \ & \ ${\rm Re}\,U^0_\Si$ & \ ${\rm Re}\,U^1_\Si$ & $\Gamma_\Si$\\
\hline
SMS LO(700)        & 7.1 & $-$16.7 & $-$1.9 & $-$10.4 & 24.7 & $-$1.5 & 1.0 & 38.3 & 21.3 \\
SMS NLO(550)       & 8.0 & $-$25.0 & $-$0.3 & $-$10.7 & 20.3 & $-$3.4 & $-$11.3 & 43.1 & 30.1 \\
SMS N$^2$LO(550)$^a$ & 7.5 & $-$24.9 & 3.4 & $-$11.0 & 20.0 & $-$4.4 & $-$10.3 & 35.0 & 34.0 \\
SMS N$^2$LO(550)$^b$ & 7.5 & $-$24.8 & 3.6 & $-$11.0 & 20.0 & $-$5.0 & $-$11.0 & 33.6 & 30.4 \\
\hline\hline
NLO13(500)         & 6.2 & $-$26.2 & 3.7 & $-$11.1 & 32.5 & $-$0.8 & 3.7 & 55.8 & 30.6 \\
NLO19(500)         & 6.0 & $-$19.5 & 3.8 & $-$10.4 & 32.7 & $-$0.8 & 11.2 & 43.2 & 22.3 \\
\hline\hline
\end{tabular}
\label{tab:Sio}
\renewcommand{\arraystretch}{1.0}
\end{table*}

\subsection{$\Si$ in symmetric nuclear matter}
\label{sec:Si}

Results for the total value and the partial-wave decomposition of the $\Sigma$
single-particle potential at $\rho_0$ are listed in Table~\ref{tab:Si}.
As one can see,
only the SMS LO interaction predicts a small repulsion, while the SMS NLO 
and N$^2$LO potentials yield an attractive $\Si$ potential at $\rho_0$.
The continuous choice results are around $10$~MeV more attractive 
compared to those for the gap choice. Also, our own
gap choice results are slightly more attractive compared to the initial
results reported in Ref.~\cite{Haidenbauer:2023qhf}. But one has to keep in
mind that the calculation in Ref.~\cite{Haidenbauer:2023qhf} is based on
a finite $\epsilon$ in the Bethe-Goldstone equation~(\ref{eq:Gmatrix}),
while now $U_\Si$ is calculated exactly by taking the 
$\epsilon\rightarrow 0$ limit.

NLO13 as well as NLO19 yield a repulsive $U_\Si$ at $\rho_0$. The primary
reason for the difference is that the contribution from the $^3S_1-^3D_1$ 
partial wave with isospin $I=3/2$ is noticeably more repulsive than that of
the SMS $YN$ potentials. This can be more clearly seen from
Table~\ref{tab:Sio}, where the isospin decomposition is listed.
A reduction of the interaction strength in that channel has become 
necessary due to constraints from recent J-PARC E40 data on $\Si^+p$ 
scattering \cite{J-PARCE40:2022nvq}. The $YN$ potentials NLO13 and NLO19
overshoot those data \cite{Haidenbauer:2023qhf}. 
In fact, other $YN$ potentials, like the FSS and fss2 interactions derived
within the constituent-quark model by Fujiwara \textit{et al.}\cite{Fujiwara:2006yh}, 
which predict a repulsive $U_\Si$ at $\rho_0$, likewise overestimate the 
cross section of the J-PARC experiment~\cite{J-PARCE40:2022nvq}. 
In contrast, $YN$ potentials like ESC16 \cite{Nagels:2015lfa} by 
the Nijmegen group that predict $\Sigma^+p$ cross sections close to those
of the SMS $YN$ potentials in the energy region of the J-PARC E40 
data (cf. Fig.~3 in that reference) 
yield likewise a slightly attractive $U_\Sigma$.
In this context let us mention that preliminary results on the $\Si^+p$ correlation function by the ALICE Collaboration point also to a somewhat weaker
$\Si^+p$ interaction in the $^3S_1$ partial wave~\cite{ALICE:2025plu}. 

\begin{figure*}[htbp]
    \centering
    \includegraphics[width=0.99\linewidth]{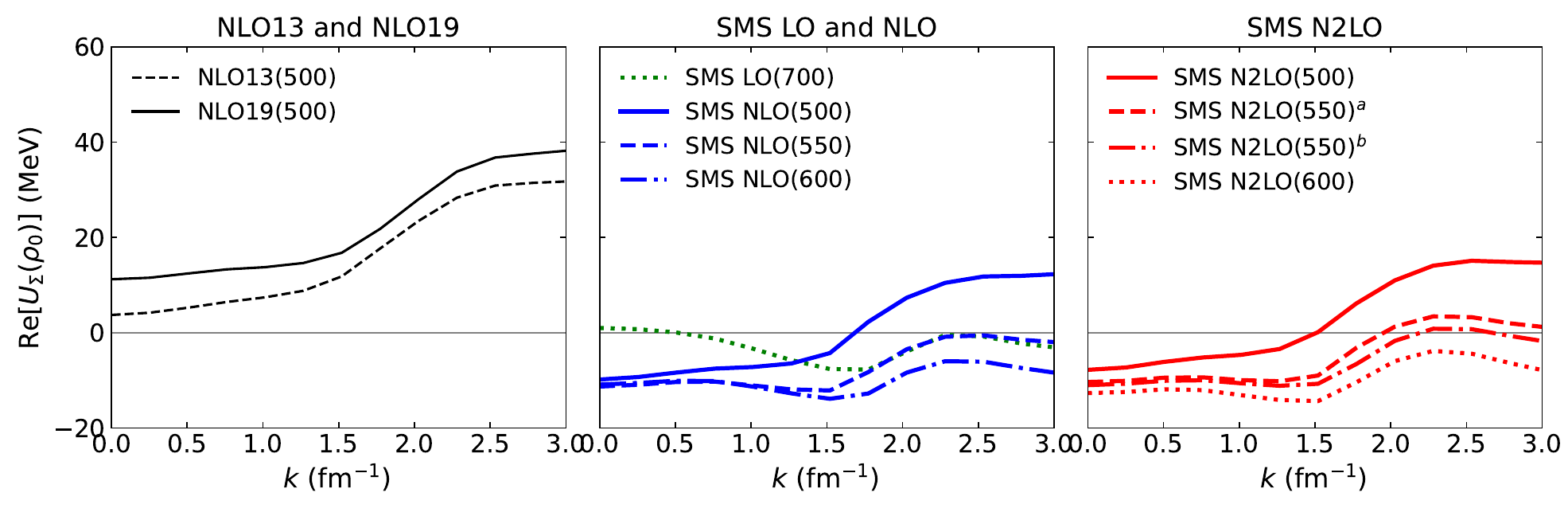}
    \includegraphics[width=0.99\linewidth]{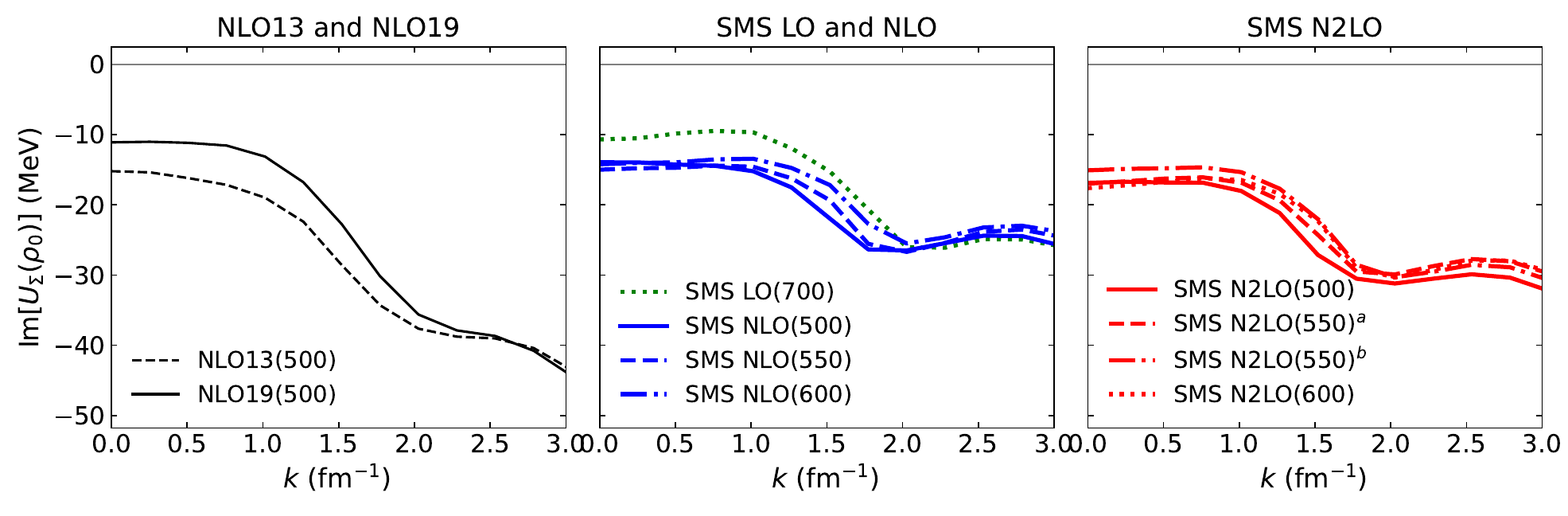}
    \caption{
    Momentum dependence of the $\Si$ single-particle potential in symmetric nuclear matter at $\rho_0$.
    Both real and imaginary parts are shown.
    }
    \label{fig:USmom_SNM}
\end{figure*}

\begin{table*}[tbhp]
\renewcommand{\arraystretch}{1.2}
\centering
\caption{
Partial-wave contributions to ${\rm Re}\,U_\Sigma(k=0)$ (in MeV) in SNM for
$\rho/\rho_0=(0.5,1.0,2.0)$, corresponding to the Fermi momenta
$k_F=(1.07,1.35,1.7)~{\rm fm}^{-1}$.
Same description of interactions as in Table \ref{tab:La}.
}
\begin{tabular}{c|c|crrrrr|r|r|r|r}
\hline\hline
& $\rho/\rho_0$ & $^1S_0$ & $^3S_1+^3D_1$ \ & $^3P_0$ & $^1P_1$ & $^3P_1$ & $^3P_2+^3F_2$ & \ $S$ waves \ & \ $P$ waves \ & \ $D$ waves \ & \, Total \, \\
\hline
SMS LO(700)     & 0.5 & $-$2.4 & 0.6 & $-$0.5 & $-$0.1 & $-$0.3 & $-$0.7 & $-$1.8 & $-$1.5 & $-$0.1 & $-$3.4 \\
     & 1.0 & $-$3.3 & 8.0 & $-$1.0 & $-$0.3 & $-$0.4 & $-$1.8 & 4.8 & $-$3.5 & $-$0.2 & 1.0 \\
     & 2.0 & $-$3.3 & 28.8 & $-$2.0 & $-$0.8 & $-$0.2 & $-$4.2 & 25.6 & $-$7.3 & $-$0.6 & 17.4 \\
\hline
SMS NLO(550)    & 0.5 & $-$2.6 & $-$5.7 & 0.2 & $-$0.2 & 0.1 & $-$1.6 & $-$8.2 & $-$1.6 & $-$0.1 & $-$9.9 \\
    & 1.0 & $-$2.7 & $-$4.6 & 0.8 & $-$0.5 & 0.6 & $-$4.5 & $-$7.3 & $-$3.6 & $-$0.2 & $-$11.3 \\
    & 2.0 & $-$1.1 & 4.8 & 3.0 & $-$0.9 & 2.4 & $-$12.1 & 3.8 & $-$7.6 & $-$0.8 & $-$4.7 \\
\hline
SMS N2LO(550)$^a$ & 0.5 & $-$3.0 & $-$6.2 & 0.1 & $-$0.4 & 0.2 & $-$0.7 & $-$9.1 & $-$0.8 & $-$0.2 & $-$10.2 \\
 & 1.0 & $-$3.4 & $-$4.9 & 0.5 & $-$0.8 & 1.0 & $-$1.6 & $-$8.2 & $-$0.9 & $-$0.9 & $-$10.3 \\
 & 2.0 & $-$2.0 & 1.8 & 2.1 & $-$1.6 & 4.0 & $-$2.1 & 0.2 & 2.4 & $-$3.5 & $-$1.5 \\
\hline
SMS N2LO(550)$^b$ & 0.5 & $-$3.0 & $-$6.1 & 0.2 & $-$1.2 & 0.3 & $-$0.3 & $-$9.1 & $-$1.0 & $-$0.2 & $-$10.4 \\
 & 1.0 & $-$3.5 & $-$4.9 & 0.8 & $-$3.3 & 1.5 & $-$0.4 & $-$8.2 & $-$1.4 & $-$0.9 & $-$11.0 \\
 & 2.0 & $-$2.1 & 3.1 & 3.1 & $-$8.2 & 5.6 & 1.0 & 1.4 & 1.5 & $-$3.6 & $-$1.1 \\
\hline\hline
NLO13(500)      & 0.5 & $-$3.7 & $-$2.4 & 0.3 & 0.0 & 0.3 & $-$0.1 & $-$6.1 & 0.5 & $-$0.1 & $-$5.6 \\
      & 1.0 & $-$4.9 & 6.4 & 1.3 & 0.2 & 1.2 & 0.2 & 1.6 & 2.8 & $-$0.2 & 3.7 \\
      & 2.0 & $-$4.3 & 43.2 & 5.2 & 1.1 & 4.2 & 2.1 & 39.2 & 12.5 & $-$0.7 & 50.9 \\
\hline
NLO19(500)      & 0.5 & $-$3.6 & 0.6 & 0.3 & 0.0 & 0.3 & $-$0.1 & $-$3.0 & 0.6 & $-$0.1 & $-$2.5 \\
      & 1.0 & $-$4.4 & 13.1 & 1.3 & 0.2 & 1.2 & 0.2 & 8.8 & 2.9 & $-$0.2 & 11.2 \\
      & 2.0 & $-$2.6 & 63.1 & 5.3 & 1.3 & 4.2 & 2.6 & 60.6 & 13.4 & $-$0.5 & 73.1 \\
\hline\hline
\end{tabular}
\label{tab:Si_rho}
\renewcommand{\arraystretch}{1.0}
\end{table*}

At the moment, it remains unclear to us how 
the constraints provided by the J-PARC data for the $\Si^+ p$ 
interaction can be reconciled with a strongly repulsive $U_\Si$ in 
the order of $10$--$50$~MeV as advocated in Ref.~\cite{Gal:2016boi}. 
Certainly, there can be additional and repulsive contributions from 
three-body forces, of similar magnitude as expected for $U_\La$.
Indeed, a corresponding calculation, based on the $\La NN$ and $\Si NN$
three-body forces introduced in Ref.~\cite{Gerstung:2020ktv} as a possible
solution of the hyperon puzzle, 
suggests that those contributions are repulsive and could be in the order of $10$ to $20$ MeV~\cite{Jinno:2025mos}. Then
one would come already close to or would be even in agreement with 
the values inferred from phenomenological potentials. 
On the other hand, the large spread in the values deduced 
from various phenomenological analyses of data on $\Si^-$ atoms 
and $(\pi^-, K^+)\Si$
spectra~\cite{,Mares:1995bm,Dabrowski:1999jy,Harada:2006yj,Kohno:2006iq,Harada:2023otu}
discloses that there is a sizable model dependence in such studies.
Therefore, one might question whether those potential parameters 
should be straightforwardly compared with results for $U_\Si$ from
a $G$-matrix calculation. After all, due to the complicated 
spin-isospin structure of the $\Si N$ interaction, where some 
of the relevant $S$ waves are attractive and others repulsive,
and the coupling to $\La N$, the situation is rather complex and
it remains unclear to what extent that can be captured by a 
simple Woods-Saxon form.

For completeness
we include in Table~\ref{tab:Sio} the isoscalar and vector decomposition 
of $U_\Sigma$~\cite{Gal:2016boi}:
\begin{align}
    \label{eq:US_isospin}
    U_{\Sigma} = \left(U^{\Sigma}_{0} + \dfrac{1}{A} \,
    U^{\Sigma}_{1} \ \bm{T}_{A} \cdot \bm{t}_{\Sigma} \right),
\end{align}
where $\bm{t}_\Sigma$ is the $\Sigma$ isospin operator and $\bm{T}_A$
is the nuclear isospin operator with $z$ projection $(Z-N)/2$.
The isoscalar $U^\Sigma_0$ and isovector $U^\Sigma_1$ components
are evaluated by $U^0_\Sigma=U_{\frac{3}{2}} + U_{\frac{1}{2}}$ and
$U^1_\Sigma = U_{\frac{3}{2}} - 2U_{\frac{1}{2}}$.
We note that the isoscalar component is identical to the total value of $U_\Sigma$ in SNM.
The isovector part is smaller than the phenomenological value listed 
in Refs.~\cite{Dover:1984nr,Gal:2016boi}. However, also here one has to 
be cautious with a direct comparison; see the discussion above.

The partial-wave breakdown for different densities is presented 
in Table~\ref{tab:Si_rho}.
The SMS NLO and N$^2$LO potentials yield similar $S$-wave 
contributions to $U_\Si$. They differ considerably from those by NLO13 
and NLO19, due to the already mentioned difference in the $I=3/2\,$ 
${}^3 S_1 - ^3 D_1$ channel.
For $P$ waves, there are again drastic differences in the results for
individual partial waves by the different potentials. In some cases
the contributions are fairly large likes those in the $^3P_2$-$^3F_2$
by the SMS NLO potential or in the $^1P_1$ by the N$^2$LO(550)$^b$
potential. Again, like in case of $U_\La$, there are cancellations
between the individual $P$-wave contributions. Due to that, 
for the two N$^2$LO potentials the total $P$-wave contributions are 
almost the same, but they differ noticeably from those for SMS NLO
and for the NLO13 and NLO19 potentials. 
Finally, there is a dramatic increase of the
$D$-wave contribution to $U_\Si$ at $2\rho_0$ in the case of the SMS 
N$^2$LO potentials. The reasons for that are the same as already discussed 
for $U_\La$ in the preceding subsection.

Figure~\ref{fig:USmom_SNM} provides the momentum dependence of
the $\Si$ single-particle potential in SNM at $\rho_0$. The real
and imaginary part of $U_\Si$ is shown. 
All SMS potentials predict a quite moderate momentum dependence for low 
$k$ values. A more noticeable variation occurs only for momenta around
$k=1.5$~fm$^{-1}$. 
Overall the results for the SMS NLO and N$^2$LO potentials are very
similar. Also the variation with the cutoff is very similar.

\begin{figure}
    \centering
    \includegraphics[width=0.99\linewidth]{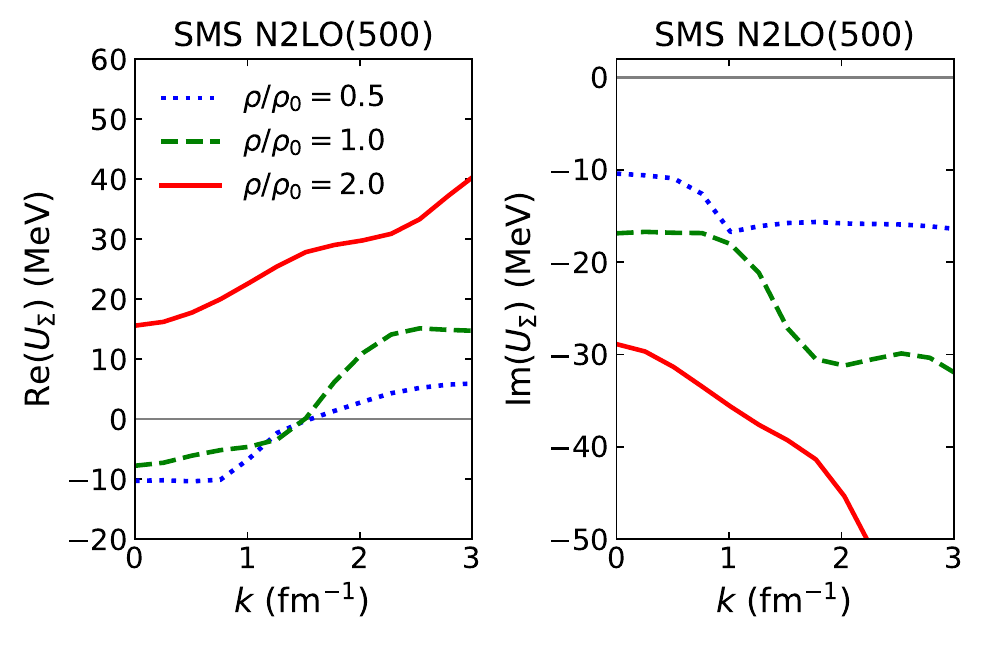}
    \caption{Momentum dependence of $U_\Si$ in symmetric nuclear matter
    for $\rho/\rho_0=(0.5,1.0,2.0)$, corresponding
    to the Fermi momenta $k_F=(1.07,1.35,1.7)~{\rm fm}^{-1}$.
    The SMS N$^2$LO(00) $YN$ potential is employed.
    }
    \label{fig:USmom_rho}
\end{figure}

Results of $U_\Si(k)$ for different densities are shown 
in Fig.~\ref{fig:USmom_rho}, exemplary for the SMS N$^2$LO(500) potential. 
Overall, the $\Si$ single-particle potential becomes
more repulsive with increasing density. Also, 
the momentum dependence near $k=0$~fm$^{-1}$ gets more
pronounced as density increases while its behavior above
$k=1$~fm$^{-1}$ varies strongly with the density.
The imaginary part becomes noticeably larger as density increases, 
indicating that $\Sigma$ hyperons are more likely to convert into
$\Lambda$ hyperons at higher density and momenta.
The imaginary part of $U_\Sigma$ is related to the conversion (or spreading)
width inside nuclear matter, $\Gamma_{\Sigma} = -2 \, {\rm Im} \,U_\Sigma(k=0)$.
Our result for SMS N$^2$LO(500), shown in Fig.~\ref{fig:USmom_rho}, 
corresponds to $\Gamma_{\Sigma}=34$ MeV at $\rho_0$.
This value is consistent with the ones reported for other $YN$ potentials~\cite{Schulze:1998jf,Kohno:2009vk}. Some results for other 
chiral $YN$ potentials are listed in Table~\ref{tab:Sio}.

\begin{figure*}[tbhp]
    \centering
    \includegraphics[width=0.99\linewidth]{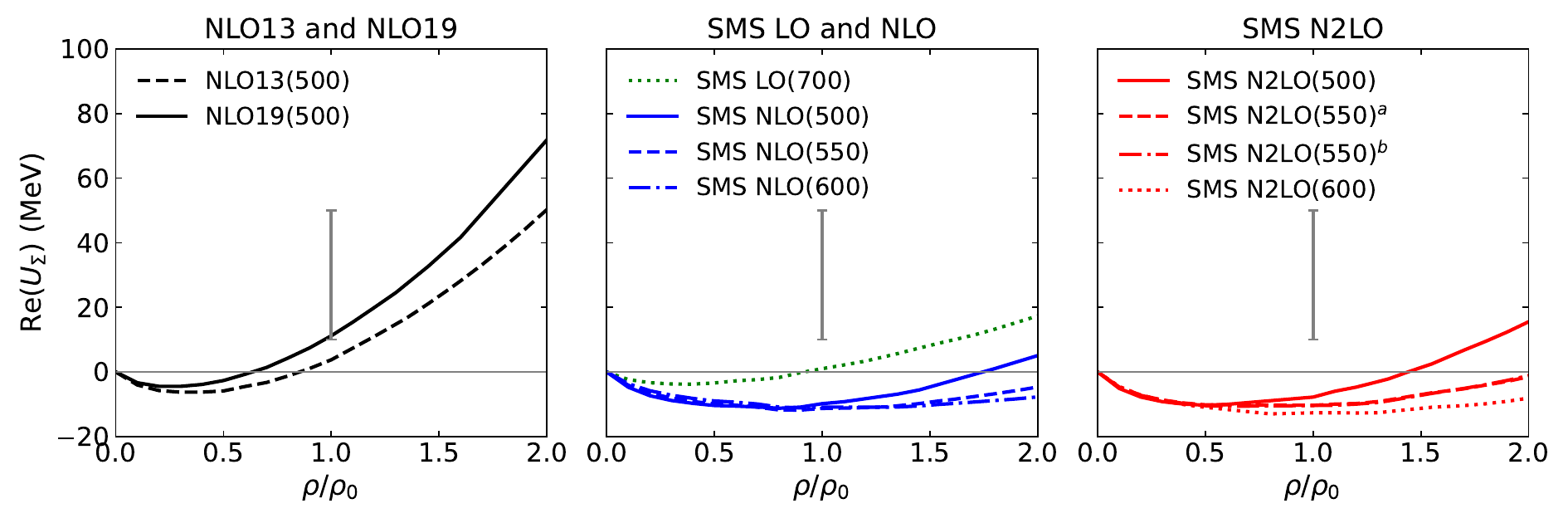}
    \caption{Density dependence of the $\Sigma$ single-particle potential in
    symmetric nuclear matter.
    The bar indicates results from phenomenological 
    analyses \cite{Gal:2016boi}.
    }
    \label{fig:US_rhoSNM}
\end{figure*}

Figure~\ref{fig:US_rhoSNM} shows the density dependence of the $\Si$
single-particle potential in SNM. 
The SMS NLO and N$^2$LO potentials predict an attractive $U_\Si$ throughout,
except for the potentials with a cutoff of 500~MeV where $U_\Si$ turns 
to repulsion at around $1.5\rho_0$. 
The results for NLO13 and NLO19 are radically different. Here $U_\Si$ 
becomes repulsive already at lower densities and then the repulsion 
increases rapidly with density. 
As already discussed above, the different behavior 
is a consequence of the constraints by the J-PARC $\Si^+p$ 
data~\cite{J-PARCE40:2022nvq} on the $\Si N$ interaction in the $I=3/2$ 
channel that have been taken into account in the SMS $YN$ potentials. 

\begin{figure*}
    \centering
    \includegraphics[width=\linewidth]{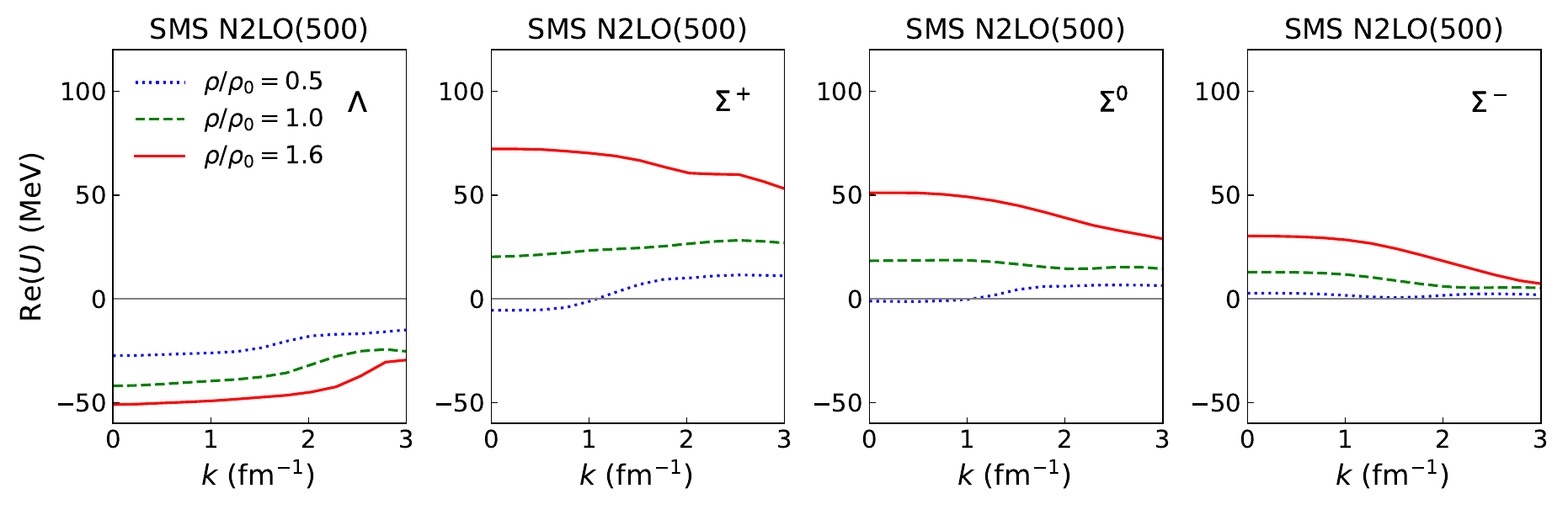}
    \caption{Momentum dependence of the real part of
    the single-particle potentials for hyperons
    in pure neutron matter for different densities
    $\rho=(0.5,1.0,1.6)$, corresponding to the Fermi momenta
    $k_F=(1.35,1.7,2.0)~{\rm fm}^{-1}$.
    }
    \label{fig:UY_rho_PNM}
\end{figure*}

\subsection{$\La$ and $\Si$ in pure neutron matter}

The dependence of $U_Y$ on the momentum for PNM is presented in Fig.~\ref{fig:UY_rho_PNM}, for $\La$, $\Si^+$, $\Si^0$, and $\Si^-$, 
using the SMS N$^2$LO(500) potential.
For all cases, $U_Y$ exhibits a fairly weak momentum dependence, 
similar to that observed in the SNM case. 
Remarkably, $U_{\Sigma^{+}}$ is larger than $U_{\Sigma^{-}}$, for
$\rho \ge \rho_0$, in contrast to the NLO13 results~\cite{Petschauer:2015nea,Kohno:2018gby}.
This is the consequence of the different isospin dependence
of the SMS $\Sigma N$ potential, i.e., the weaker $I=3/2$ repulsion 
imposed by the J-PARC E40 data~\cite{J-PARCE40:2022nvq}, 
as already seen in the SNM calculation; cf. Table~\ref{tab:Sio}.

\begin{figure*}[tbhp]
    \centering
    \includegraphics[width=0.99\linewidth]{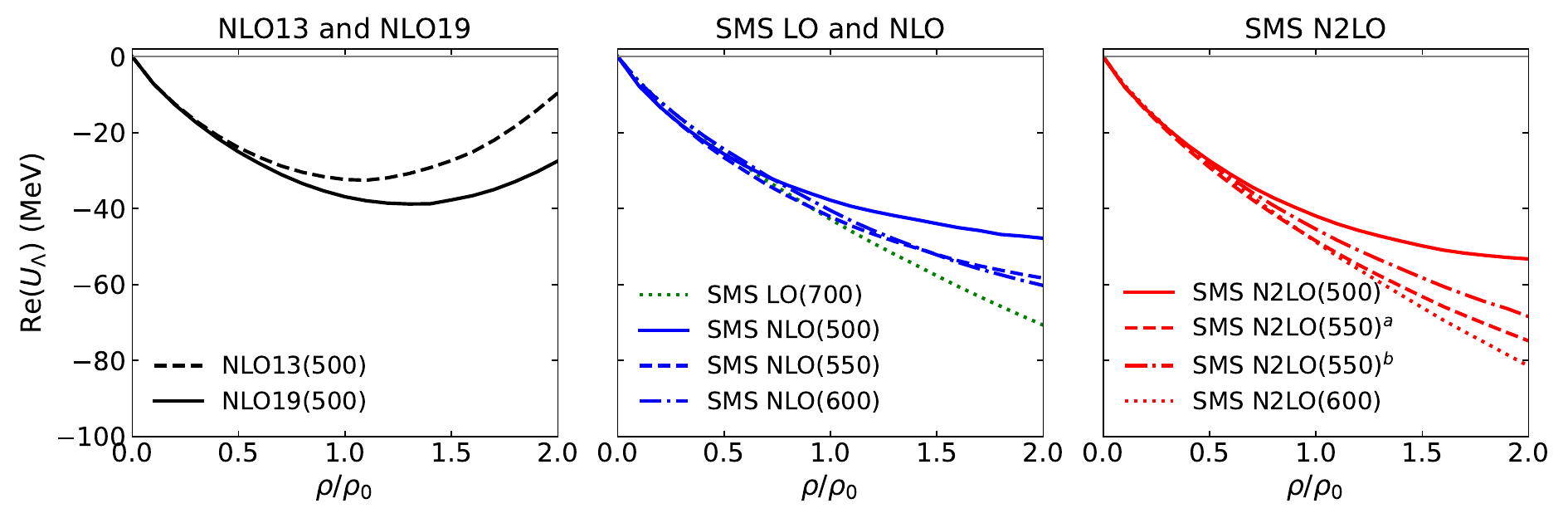}
    \includegraphics[width=0.99\linewidth]{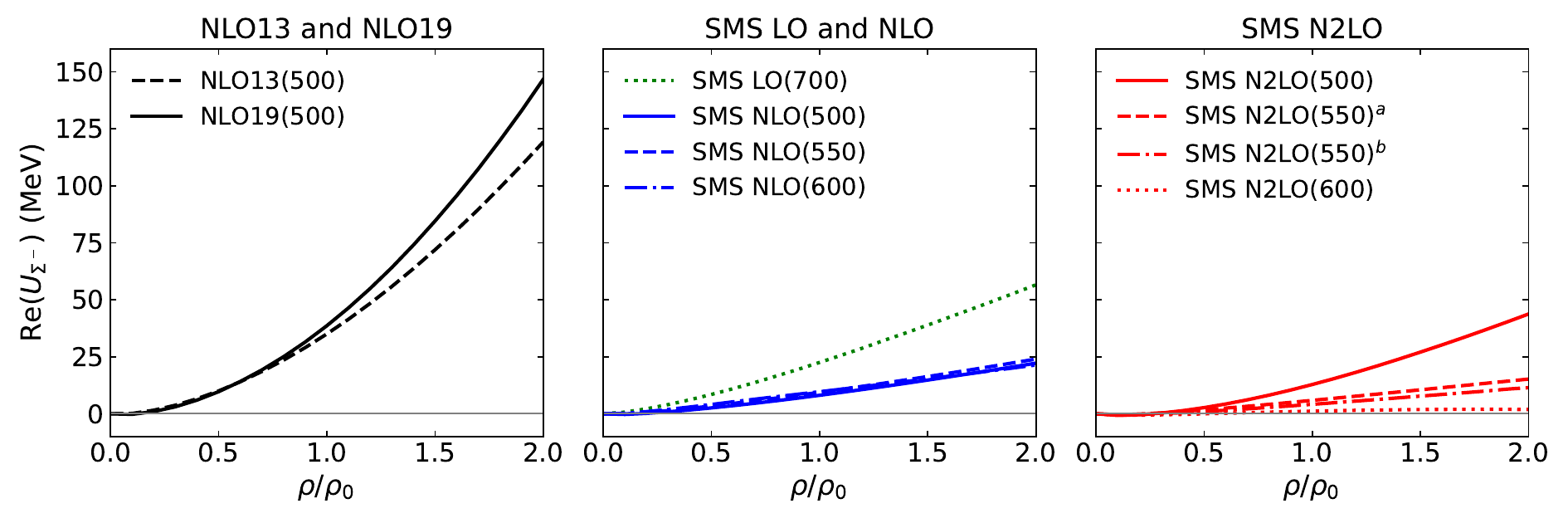}
    \caption{Density dependence of the $\Lambda$ (upper panels) and $\Sigma^-$ (lower panels) single-particle
    potentials in pure neutron matter.
    }
    \label{fig:UY_rhoPNM}
\end{figure*}

In Fig.~\ref{fig:UY_rhoPNM} the density dependence of the hyperon 
single-particle potentials in PNM is presented. With regard to the
$\La$, the SMS NLO and N$^2$LO results exhibit a steady increase of
the attraction with density,
whereas those from NLO13 and NLO19 become noticeably less attractive
from around $\rho_0$ onward. Nevertheless the difference between the SNM
and PNM results are small for the $\La$, since it is an isoscalar
particle. 

In contrast, the isospin $I=1$ hyperon $\Sigma$ is strongly affected by the
relative fraction of protons and neutrons in nuclear matter. 
The $\Sigma^- n$ system has isospin $I=3/2$. The corresponding interaction 
is identical to the one for $\Si^+p$ when isospin-breaking effects are
disregarded and predominantly repulsive. However, in the case of the SMS $YN$ 
potentials the repulsion is significantly reduced in comparison to 
NLO13 and NLO19, due to their adjustment to the new $\Si^+ p$ data, as
already mentioned. As a result, the corresponding $\Sigma^-$ single-particle
potential is not strongly repulsive anymore.  

\section{Uncertainty estimate}
\label{sec:uncertaity}

In this section, we provide an uncertainty estimate for
the hyperon single-particle potentials at $k=0$, i.e., for the quantity
that is associated with the binding energy of hyperons in infinite
nuclear matter. In particular, we estimate and discuss the truncation 
error in the chiral expansion, following the method proposed by Epelbaum,
Krebs, and Mei\ss ner (EKM) \cite{Epelbaum:2014efa,LENPIC:2015qsz}. 
It combines information about the expected size and actual size of 
higher-order corrections. 
It can be applied to any observable $X$ which has been evaluated up to
a specific order $i$ in the chiral expansion, $X^{(i)}$.
The concrete expressions for the corresponding uncertainty $\delta X^{(i)}$
are~\cite{Epelbaum:2014sza}
\begin{eqnarray}
\label{eq:EKM}
	\delta X^{\text{LO}} &=& Q^2 \left|X^{\text{LO}}\right|, \nonumber\\
	\delta X^{\text{NLO}} &=&\max\Big(
		Q^3 \left|X^{\text{LO}}\right|,
		Q\left|X^{\text{NLO}}-X^{\text{LO}}\right|
	\Big), \nonumber\\
	\delta X^{\text{N$^2$LO}} &=& \max\Big(
		Q^4 \left|X^{\text{LO}}\right|,
		Q^2 \left|X^{\text{NLO}} - X^{\text{LO}}\right|, \nonumber \\
		&&\qquad\quad Q \left|X^{\text{N$^2$LO}}-X^{\text{NLO}}\right|
	\Big) \ , 
\end{eqnarray}
with the additional constraint for the theoretical uncertainties at LO
and NLO to have at least the size of the actual higher-order
contributions \cite{Epelbaum:2014efa}. 
Here, $Q$ represents the expansion scale in the chiral expansion, 
which is given by $Q\in \{p/\Lambda_b,M_\pi/\Lambda_b\}$,
where $p$ is the typical momentum of the baryons, $M_\pi$ 
is the pion mass, and $\Lambda_b$ is the breakdown scale of the 
chiral expansion.
This method has been already applied to nuclear matter properties 
in the work by Hu \textit{et al.}\cite{Hu:2016nkw}.
Thereby, the momentum scale $p$ has been identified with the Fermi momentum
$k_F$. We adopt the same prescription in our work. 

For the breakdown scale we use $\Lambda_b = 480$ and $600$~MeV
for SNM and PNM, respectively. This choice was guided by 
the Bayesian analysis of nuclear matter properties by 
Hu \textit{et al.}\cite{Hu:2019zwa}, which suggested a 
breakdown scale of $480$~MeV for SNM and $660$--$720$~MeV for PNM.
In the initial study by Hu \textit{et al.}\cite{Hu:2016nkw} with the
EKM method, $\Lambda_b = 600$~MeV has been used.
The breakdown scale in the original work by EKM was estimated 
to be around $\Lambda_b = 400$--$600$~MeV~\cite{Epelbaum:2014sza}, based 
on results for $NN$ phase shifts and scattering observables.  
The typical values of $Q$ are $0.55$ and $0.69$ at $\rho_0$ and 
$2\rho_0$ in SNM, respectively.
For comparison, in the study of the uncertainty for light $\Lambda$
hypernuclei a scale of $Q=0.4$ was found appropriate \cite{Le:2023bfj}.
As pointed out in Ref.~\cite{Epelbaum:2014sza}, such a simple 
estimation of the theoretical uncertainty does not provide a statistical
interpretation. Nonetheless, the procedure can be interpreted in 
a Bayesian sense~\cite{Furnstahl:2015rha}.

\begin{figure*}
    \centering
    \includegraphics[width=0.49\linewidth]{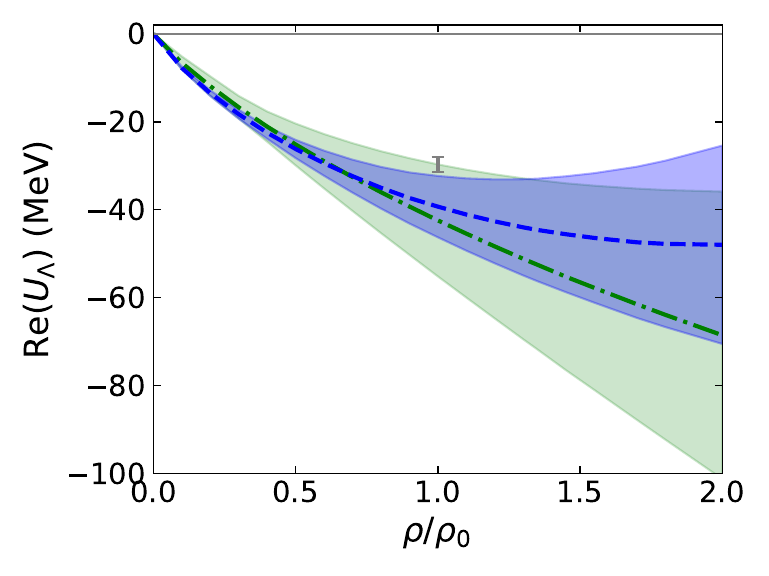}
    \includegraphics[width=0.49\linewidth]{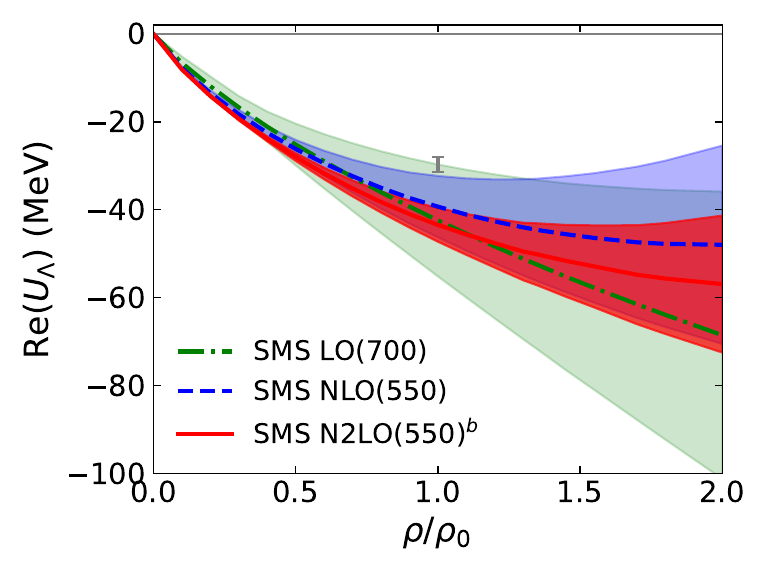}
    \caption{Uncertainty estimate of the $\Lambda$ single-particle potentials in
    symmetric nuclear matter.
    The SMS NLO and N$^2$LO interactions with cutoff $550$~MeV are employed, while for SMS LO
    the potential with $700$~MeV cutoff is used.
    The bar symbolizes the quasiempirical value~\cite{Gal:2016boi}. 
    }
    \label{fig:ULSNM550_uncertainty}
\end{figure*}
\begin{figure*}
    \centering
    \includegraphics[width=0.49\linewidth]{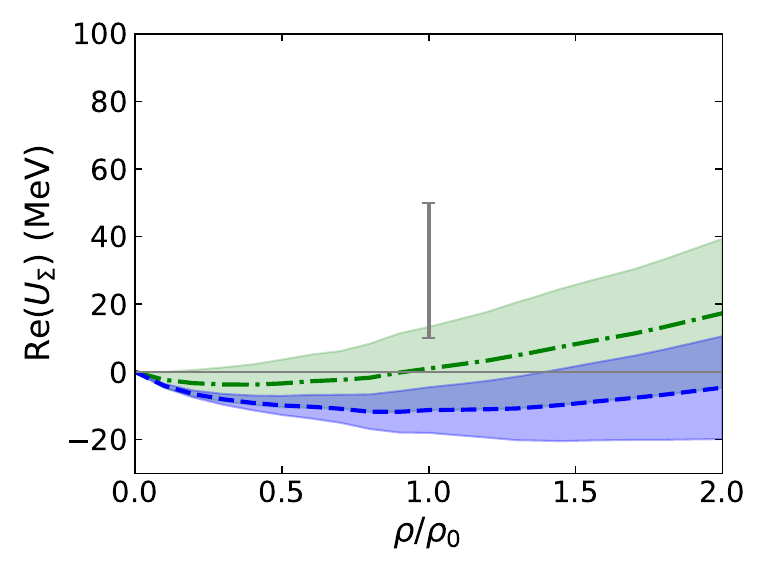}
    \includegraphics[width=0.49\linewidth]{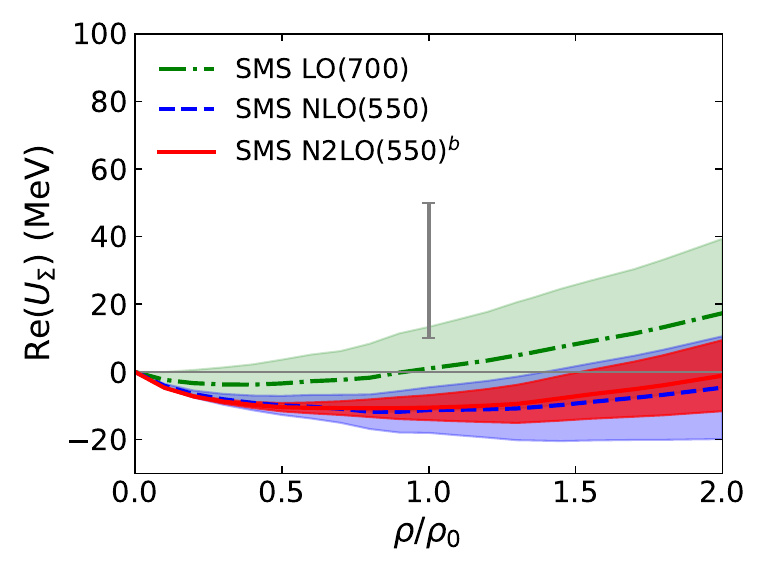}
    \caption{Same as Fig.~\ref{fig:ULSNM550_uncertainty},  but for $\Sigma$.
    The bar indicates results from phenomenological 
    analyses, taken from Ref.~\cite{Gal:2016boi}.
    }
    \label{fig:USSNM550_uncertainty}
\end{figure*}

\begin{figure*}
    \centering
    \includegraphics[width=0.49\linewidth]{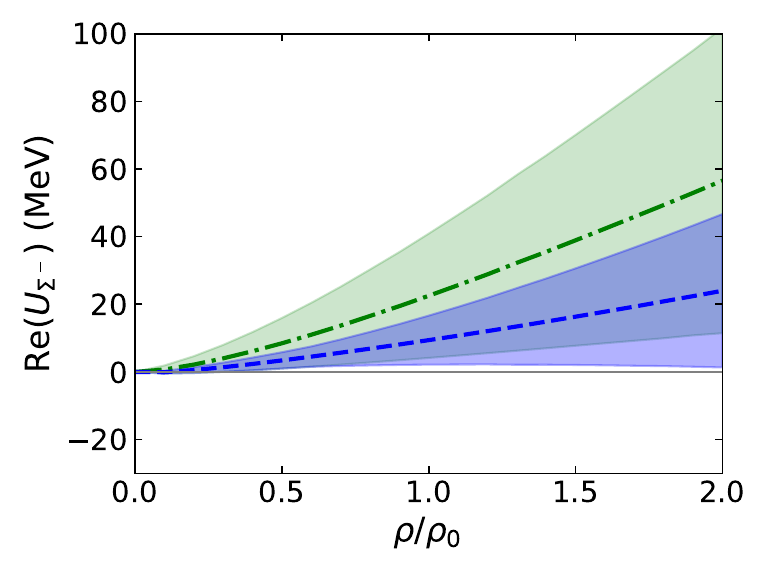}
    \includegraphics[width=0.49\linewidth]{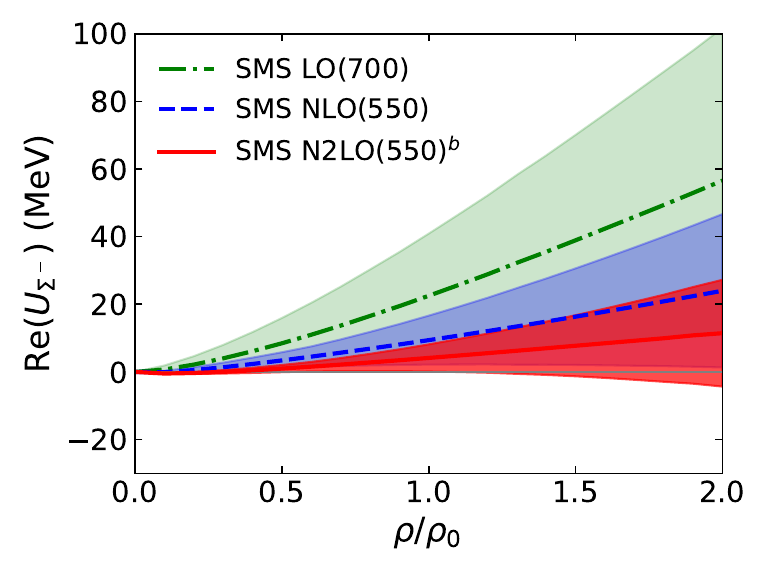}
    \caption{Uncertainty estimate of the $\Sigma^-$ single-particle potentials in
    pure neutron matter.
    }
    \label{fig:USmPNM550_uncertainty}
\end{figure*}

In Fig.~\ref{fig:ULSNM550_uncertainty}, the uncertainty estimate
for the density dependence of $U_\La (k=0)$ in SNM is shown.
We selected the SMS LO(700), SMS NLO(550),
and SMS N$^2$LO(550)$^b$ $YN$ potentials as an exemplary set.
The other cutoffs of $500$ and $600$~MeV result 
in very similar uncertainty bands for the SNM calculations.
The upper limits of the LO and NLO uncertainty bands at
$\rho_0$ touch the 
quasiempirical value of $U_\Lambda\approx -30$~MeV. 
In contrast, the N$^2$LO band underestimates it.
The actual estimates at $\rho_0$ are
\begin{align}
    U^{\rm NLO}_\Lambda(\rho_0) &= -39.3 \pm 7.0~\text{MeV}, \nonumber \\
    U^{\rm N^2LO}_\Lambda(\rho_0) &= -43.5 \pm 3.8~\text{MeV}.
    \label{eq:UL_uncertainty}
\end{align}
Obviously, the uncertainty is of an order which brings the SMS NLO results
(and those of the NLO13 and NLO19 $YN$ potentials) practically in line 
with the quasiempirical value of $U_\La$. An interesting aspect is, of 
course, that the uncertainty of the NLO result provides also a glimpse 
on the size of missing $YNN$ three-body forces. 
The present estimate is compatible with the actual contribution 
of an effective three-body force evaluated in Ref.~\cite{Gerstung:2020ktv},
established within the assumption of decuplet saturation and based on 
LECs constrained from dimensional scaling arguments. It is also comparable 
to the value suggested in Ref.~\cite{Friedman:2023ucs}
within a phenomenological analysis. 
Comparing the uncertainty estimate with the NLO results for different cutoffs
in Table~\ref{tab:La}, one can see that the cutoff variations are noticeably
smaller. This confirms again the conclusion that the cutoff variation does 
not provide a reliable measure for estimating uncertainties,
stressed already in earlier studies of nuclear matter
\cite{Hu:2016nkw,Sammarruca:2014zia}. 
Indeed, as emphasized in Ref.~\cite{Epelbaum:2014sza}, in general the 
residual cutoff dependence underestimates the theoretical uncertainty. 
Since the calculation at N$^2$LO is incomplete, i.e., three-body forces are 
missing, one has to be somewhat cautious with the interpretation 
of the corresponding result. Strictly speaking, the actual uncertainty 
is still the one obtained for the result at NLO. This concerns not only
the estimate at $\rho_0$ but also the one at the highest density 
considered by us. As one can see from Fig.~\ref{fig:ULSNM550_uncertainty},
the uncertainty of $U_\La$ increases rapidly with density and amounts 
to roughly $\pm 20$~MeV at $2\rho_0$.

Additional uncertainties arise due to the issues already discussed in 
Sec.~\ref{sec:La}. Since the $P$-wave interactions in the $\La N$
system are not directly constrained, only indirectly via $\Si^-p$
and $\Si^+p$ differential cross sections and the assumed SU(3) symmetry, 
their contribution to $U_\La$ at $\rho_0$ varies in the order of $4$~MeV 
for the considered chiral $YN$ potentials, cf. Table~\ref{tab:La_rho}, 
and by almost $10$~MeV at $2\rho_0$. 
In the case of the N$^2$LO calculation even uncertainties in the $D$ waves 
might become relevant.
Actually, in the context of the N$^2$LO potential it is instructive to take 
a look at the nuclear matter case, where calculations have been performed 
up to N$^4$LO. There the result at N$^2$LO sticks out from the general trend 
as being particularly attractive; see Fig.~1 in Ref.~\cite{Hu:2016nkw}.
Thus one can speculate that also for the $\Lambda$ the N$^2$LO result
is somewhat exceptional and a fully converged calculation would rather
yield values for $U_\La$ that are close to those obtained for the SMS
NLO $YN$ potentials. 

Note that other uncertainties, like those associated with the use of the 
BHF approximation~\cite{Song:1998zz,Baldo:2001mv,Lu:2017nbi,Shang:2021msd}, are not considered here.

The uncertainty estimation for $\Sigma$ in SNM is presented in Fig.~\ref{fig:USSNM550_uncertainty}.
The corresponding uncertainty at $\rho_0$ is
\begin{align}
    U^{\rm NLO}_\Sigma(\rho_0) &= -11.3 \pm 6.7~\text{MeV},
    \nonumber\\
    U^{\rm N^2LO}_\Sigma(\rho_0) &= -10.6 \pm 3.7~\text{MeV}
\end{align}
The uncertainty bands for the NLO and N$^2$LO potentials are similar because 
there is not much difference between the results at NLO and N$^2$LO, as
can be seen from Fig.~\ref{fig:USmom_SNM}. Obviously, $U_\Si$ is
predicted to be attractive up to $1.5\rho_0$, even when considering the
theoretical uncertainty.

In Fig.~\ref{fig:USmPNM550_uncertainty} we show the $\Sigma^-$
single-particle potential in PNM. The values at $\rho_0$ in PNM are
\begin{align}
    U^{\rm NLO}_{\Sigma^{-}}(\rho_0) &= 9.4 \pm 7.2~\text{MeV},
    \nonumber\\
    U^{\rm N^2LO}_{\Sigma^{-}}(\rho_0) &= 4.1 \pm 4.0~\text{MeV}.
\end{align}
Compared to the case of $U_\Sigma$ in SNM, the N$^2$LO uncertainty band
is noticeably larger due to the fact that for a given $\rho_0$ the Fermi momentum in PNM is larger. 
Obviously $U_{\Si^-}$ is predominantly repulsive, for the SMS NLO as well
as for the N$^2$LO potentials. 

We do not show results for $\La$ in PNM since the uncertainties are
very similar to those in SNM, reflecting the isoscalar nature of the 
$\La$ hyperon.

We have made some exploratory calculations utilizing a Bayesian approach~\cite{Melendez:2019izc}.
Those yield results very similar to the EKM method. Therefore, we postpone
a full-fledged Bayesian analysis for the future.

\section{Summary and outlook}
\label{sec:summary}

We have investigated the properties of hyperons in
nuclear matter within
the self-consistent Brueckner-Hartree-Fock approach
using the continuous choice for intermediate states.
We employed the recently established semilocal
momentum-space regularized (SMS) hyperon-nucleon
interaction up to next-to-next-to-leading order,
enabling a systematic investigation of the order-by-order convergence.
We also used the $NN$ interaction by Reinert \textit{et al.}\cite{Reinert:2017usi}, 
regularized in the same scheme as the hyperon-nucleon potential, 
to evaluate the nucleon single-particle potential.

The $\La$ single-particle potentials from the SMS $YN$ interactions turned 
out to be qualitatively similar to those predicted by the NLO13 and NLO19
interactions, at least up to nuclear matter saturation density. 
However, overall, the resulting in-medium potentials from the new
$YN$ interactions are slightly more attractive, in particular those 
from the N$^2$LO interaction. This concerns first of all the value
of $U_\La$ at the momentum $k=0$~MeV, associated with the binding energy
of the $\La$ in infinite nuclear matter, but is reflected also in the 
momentum dependence. 
For the $\Si$ case the results for the SMS $YN$ potentials differ 
noticeably from those for our earlier chiral $YN$ potentials. 
Specifically, now the single-particle potential in symmetric nuclear 
matter at the saturation density is even attractive, $U_\Si\approx -10$~MeV,
while results in the order of $U_\Si \approx +10$~MeV are predicted by the
NLO13/NLO19 potentials. 
The reason for the difference is new constraints
from $\Si^+ p$ amd $\Si^-p$ differential cross section data from the  
J-PARC E40 experiment~\cite{J-PARCE40:2021bgw,J-PARCE40:2021qxa,J-PARCE40:2022nvq}, which have been taken into account when the
SMS $YN$ potentials have been established. 

In the present work we reported also on the first effort 
to estimate the error due to the truncation in the
chiral expansion for the single-particle potentials.
Thereby we followed a procedure proposed by Epelbaum, 
Krebs, and Mei{\ss}ner~\cite{Epelbaum:2014sza}, a simple but efficient
method which combines information about the expected size and the 
actual size of higher-order corrections.
The estimated uncertainty relevant for the $\La$ is in the order 
of $7$~MeV at nuclear matter saturation density. It is comparable with 
the difference of the theoretical results for $U_\La$ at NLO to the
usually quoted quasiempirical value of $U_\La\approx -30$~MeV.  

The present results at N$^2$LO are incomplete because corresponding
contributions from leading three-body forces, which arise 
at that order of the chiral expansion, have not been taken into 
account. Indeed, the uncertainty estimate for the NLO result 
allows a first rough conclusion on the actual size of possible
contributions from $YNN$ three-body forces. An explicit calculation
of their effect is, however, extremely challenging. 
One needs to solve the Bethe-Faddeev equation
\cite{Bethe:1965zz,Day:1966zza} with chiral three-body forces. 
So far pertinent investigations in the context of chiral $NN$ 
\cite{Sammarruca:2014zia,Sammarruca:2018va}
and $YN$ \cite{Haidenbauer:2016vfq,Gerstung:2020ktv}
potentials have been limited to an approximation, namely to the 
application of density-dependent two-body potentials derived from the 
N$^2$LO three-body force. 
Such an effective interaction can be obtained by summing 
one particle over the occupied states in the Fermi sea 
\cite{Holt:2009uk,Holt:2009ty,Holt:2019bah,Petschauer:2016pbn}. 
In the on-shell approximation the resulting $YN$ ($NN$) interaction 
can be expressed in analytical form with operator structures 
identical to those of free-space $YN$ ($NN$) interactions; see 
Ref.~\cite{Petschauer:2016pbn} for details. This idea has been already 
exploited by Gerstung \textit{et al.}\cite{Gerstung:2020ktv} and they showed that
with such an effective three-body force one can not only reproduce 
the quasiempirical value of $U_\La$ at $\rho_0$ but even
solve the hyperon puzzle. 
In principle, one could avoid the on-shell approximation
by summing one particle over the occupied states in
the Fermi sea numerically.
In this case, there would be a much more direct connection between the 
three-body forces as used in {\it ab initio} few-body 
calculations \cite{Le:2024rkd} 
and the ones that enter the $G$-matrix calculation.
Work along this line is left for the future.

\vskip 0.4cm
\begin{acknowledgements}
We would like to thank Jinniu Hu and Isaac Vida\~na for sharing their 
results and for fruitful communication, Thomas Duguet, Avraham Gal,
Andreas Nogga, and Eulogio Oset for useful discussions,
and Dominik Gerstung for kindly sharing his code with us.
This work was supported in part by JST SPRING (Grant No. JPMJSP2110) and
by the Grants-in Aid for Scientific Research from JSPS (Grant No. JP25KJ1584), by the 
European Research Council (ERC) under the European Union’s Horizon 2020 research and innovation programme (ERC AdG EXOTIC, Grant Agreement No. 101018170) and the  CAS President’s International Fellowship Initiative (PIFI) (Grant No. 2025PD0022).
\end{acknowledgements}

\label{Bibliography}
\bibliographystyle{unsrturl}
\bibliography{bib2.bib,bibliography.bib}

\begin{thebibliography}{100}

\bibitem{Haidenbauer:2023qhf}
J.~Haidenbauer, U.-G. Mei\ss{}ner, A.~Nogga, and H.~Le.
\newblock {Hyperon\textendash{}nucleon interaction in chiral effective field theory at next-to-next-to-leading order}.
\newblock {\em Eur. Phys. J. A}, 59(3):63, 2023.
\newblock \href {https://arxiv.org/abs/2301.00722} {\path{arXiv:2301.00722}}, \href {https://doi.org/10.1140/epja/s10050-023-00960-6} {\path{doi:10.1140/epja/s10050-023-00960-6}}.

\bibitem{Miyagawa:1993rd}
K.~Miyagawa and W.~Gl{\"o}ckle.
\newblock {Hypertriton calculation with meson theoretical nucleon-nucleon and hyperon nucleon interactions}.
\newblock {\em Phys. Rev. C}, 48:2576, 1993.
\newblock \href {https://doi.org/10.1103/PhysRevC.48.2576} {\path{doi:10.1103/PhysRevC.48.2576}}.

\bibitem{Nogga:2001ef}
A.~Nogga, H.~Kamada, and W.~Gl{\"o}ckle.
\newblock {The Hypernuclei $^4_\Lambda$He and $^4_{\Lambda}$He: Challenges for modern hyperon nucleon forces}.
\newblock {\em Phys. Rev. Lett.}, 88:172501, 2002.
\newblock \href {https://arxiv.org/abs/nucl-th/0112060} {\path{arXiv:nucl-th/0112060}}, \href {https://doi.org/10.1103/PhysRevLett.88.172501} {\path{doi:10.1103/PhysRevLett.88.172501}}.

\bibitem{Nogga:2013pwa}
A.~Nogga.
\newblock {Light hypernuclei based on chiral and phenomenological interactions}.
\newblock {\em Nucl. Phys. A}, 914:140--150, 2013.
\newblock \href {https://doi.org/10.1016/j.nuclphysa.2013.02.053} {\path{doi:10.1016/j.nuclphysa.2013.02.053}}.

\bibitem{Haidenbauer:2021wld}
J.~Haidenbauer, U.-G. Mei\ss{}ner, and A.~Nogga.
\newblock {Constraints on the $\varLambda $-Neutron Interaction from Charge Symmetry Breaking in the $\mathbf {^4_\Lambda \mathrm{He}}$ - $\mathbf {^4_\Lambda \mathrm{H}}$ Hypernuclei}.
\newblock {\em Few Body Syst.}, 62(4):105, 2021.
\newblock \href {https://arxiv.org/abs/2107.01134} {\path{arXiv:2107.01134}}, \href {https://doi.org/10.1007/s00601-021-01684-3} {\path{doi:10.1007/s00601-021-01684-3}}.

\bibitem{Wirth:2014apa}
R.~Wirth, P.~Gazda, D.and~Navr{\'a}til, A.~Calci, J.~Langhammer, and R.~Roth.
\newblock {Ab Initio Description of p-Shell Hypernuclei}.
\newblock {\em Phys. Rev. Lett.}, 113(19):192502, 2014.
\newblock \href {https://arxiv.org/abs/1403.3067} {\path{arXiv:1403.3067}}, \href {https://doi.org/10.1103/PhysRevLett.113.192502} {\path{doi:10.1103/PhysRevLett.113.192502}}.

\bibitem{Wirth:2017lso}
R.~Wirth and R.~Roth.
\newblock {Light Neutron-Rich Hypernuclei from the Importance-Truncated No-Core Shell Model}.
\newblock {\em Phys. Lett. B}, 779:336--341, 2018.
\newblock \href {https://arxiv.org/abs/1710.04880} {\path{arXiv:1710.04880}}, \href {https://doi.org/10.1016/j.physletb.2018.02.021} {\path{doi:10.1016/j.physletb.2018.02.021}}.

\bibitem{Le:2020zdu}
H.~Le, J.~Haidenbauer, U.-G. Mei\ss{}ner, and A.~Nogga.
\newblock {Jacobi no-core shell model for $p$-shell hypernuclei}.
\newblock {\em Eur. Phys. J. A}, 56(12):301, 2020.
\newblock \href {https://arxiv.org/abs/2008.11565} {\path{arXiv:2008.11565}}, \href {https://doi.org/10.1140/epja/s10050-020-00314-6} {\path{doi:10.1140/epja/s10050-020-00314-6}}.

\bibitem{Le:2022ikc}
H.~Le, J.~Haidenbauer, U.-G. Mei\ss{}ner, and A.~Nogga.
\newblock {Ab initio calculation of charge-symmetry breaking in A=7 and 8 \ensuremath{\Lambda} hypernuclei}.
\newblock {\em Phys. Rev. C}, 107(2):024002, 2023.
\newblock \href {https://arxiv.org/abs/2210.03387} {\path{arXiv:2210.03387}}, \href {https://doi.org/10.1103/PhysRevC.107.024002} {\path{doi:10.1103/PhysRevC.107.024002}}.

\bibitem{Le:2024aox}
Hoai Le, Johann Haidenbauer, Hiroyuki Kamada, Michio Kohno, Ulf-G. Mei{\ss}ner, Kazuya Miyagawa, and Andreas Nogga.
\newblock {Benchmarking $\mathbf {\Lambda }$NN three-body forces and first predictions for $A=3-5$ hypernuclei}.
\newblock {\em Eur. Phys. J. A}, 61(2):21, 2025.
\newblock \href {https://arxiv.org/abs/2407.02064} {\path{arXiv:2407.02064}}, \href {https://doi.org/10.1140/epja/s10050-024-01474-5} {\path{doi:10.1140/epja/s10050-024-01474-5}}.

\bibitem{Hildenbrand:2024ypw}
Fabian Hildenbrand, Serdar Elhatisari, Zhengxue Ren, and Ulf-G. Mei{\ss}ner.
\newblock {Towards hypernuclei from nuclear lattice effective field theory}.
\newblock {\em Eur. Phys. J. A}, 60(10):215, 2024.
\newblock \href {https://arxiv.org/abs/2406.17638} {\path{arXiv:2406.17638}}, \href {https://doi.org/10.1140/epja/s10050-024-01427-y} {\path{doi:10.1140/epja/s10050-024-01427-y}}.

\bibitem{Yamamoto:1994tc}
Y.~Yamamoto, T.~Motoba, H.~Himeno, K.~Ikeda, and S.~Nagata.
\newblock {Hyperon nucleon and hyperon-hyperon interactions in nuclei}.
\newblock {\em Prog. Theor. Phys. Suppl.}, 117:361--389, 1994.
\newblock \href {https://doi.org/10.1143/PTPS.117.361} {\path{doi:10.1143/PTPS.117.361}}.

\bibitem{Hiyama:1997ub}
E.~Hiyama, M.~Kamimura, T.~Motoba, T.~Yamada, and Y.~Yamamoto.
\newblock {Three- and four-body cluster models of hypernuclei using the G-matrix Lambda N interaction: (Lambda)Be-9, (Lambda)C-13, (Lambda Lambda)He-6 and (Lambda Lambda)Be-10}.
\newblock {\em Prog. Theor. Phys.}, 97:881--899, 1997.
\newblock \href {https://doi.org/10.1143/PTP.97.881} {\path{doi:10.1143/PTP.97.881}}.

\bibitem{Vidana:1998ed}
I.~Vida\~na, A.~Polls, A.~Ramos, and M.~Hjorth-Jensen.
\newblock {Hyperon properties in finite nuclei using realistic YN interactions}.
\newblock {\em Nucl. Phys. A}, 644:201--220, 1998.
\newblock \href {https://arxiv.org/abs/nucl-th/9805032} {\path{arXiv:nucl-th/9805032}}, \href {https://doi.org/10.1016/S0375-9474(98)00599-5} {\path{doi:10.1016/S0375-9474(98)00599-5}}.

\bibitem{Fujiwara:2006fj}
Y.~Fujiwara, M.~Kohno, and Y.~Suzuki.
\newblock {Lambda alpha, Sigma alpha and Xi alpha potentials derived from the SU(6) quark-model baryon-baryon interaction}.
\newblock {\em Nucl. Phys. A}, 784:161--187, 2007.
\newblock \href {https://arxiv.org/abs/nucl-th/0610086} {\path{arXiv:nucl-th/0610086}}, \href {https://doi.org/10.1016/j.nuclphysa.2006.12.005} {\path{doi:10.1016/j.nuclphysa.2006.12.005}}.

\bibitem{Vidana:2016ayd}
Isaac Vida{\~n}a.
\newblock {Single-particle spectral function of the $\Lambda$ hyperon in finite nuclei}.
\newblock {\em Nucl. Phys. A}, 958:48--70, 2017.
\newblock \href {https://arxiv.org/abs/1603.05635} {\path{arXiv:1603.05635}}, \href {https://doi.org/10.1016/j.nuclphysa.2016.11.002} {\path{doi:10.1016/j.nuclphysa.2016.11.002}}.

\bibitem{Haidenbauer:2019thx}
Johann Haidenbauer and Isaac Vida\~na.
\newblock {Structure of single-$\Lambda$ hypernuclei with chiral hyperon-nucleon potentials}.
\newblock {\em Eur. Phys. J. A}, 56(2):55, 2020.
\newblock \href {https://arxiv.org/abs/1910.02695} {\path{arXiv:1910.02695}}, \href {https://doi.org/10.1140/epja/s10050-020-00055-6} {\path{doi:10.1140/epja/s10050-020-00055-6}}.

\bibitem{Haidenbauer:2016vfq}
J.~Haidenbauer, U.-G. Mei\ss{}ner, N.~Kaiser, and W.~Weise.
\newblock {Lambda-nuclear interactions and hyperon puzzle in neutron stars}.
\newblock {\em Eur. Phys. J. A}, 53(6):121, 2017.
\newblock \href {https://arxiv.org/abs/1612.03758} {\path{arXiv:1612.03758}}, \href {https://doi.org/10.1140/epja/i2017-12316-4} {\path{doi:10.1140/epja/i2017-12316-4}}.

\bibitem{Gerstung:2020ktv}
D.~Gerstung, N.~Kaiser, and W.~Weise.
\newblock {Hyperon\textendash{}nucleon three-body forces and strangeness in neutron stars}.
\newblock {\em Eur. Phys. J. A}, 56(6):175, 2020.
\newblock \href {https://arxiv.org/abs/2001.10563} {\path{arXiv:2001.10563}}, \href {https://doi.org/10.1140/epja/s10050-020-00180-2} {\path{doi:10.1140/epja/s10050-020-00180-2}}.

\bibitem{Logoteta:2019utx}
Domenico Logoteta, Isaac Vida\~ana, and Ignazio Bombaci.
\newblock {Impact of chiral hyperonic three-body forces on neutron stars}.
\newblock {\em Eur. Phys. J. A}, 55(11):207, 2019.
\newblock \href {https://arxiv.org/abs/1906.11722} {\path{arXiv:1906.11722}}, \href {https://doi.org/10.1140/epja/i2019-12909-9} {\path{doi:10.1140/epja/i2019-12909-9}}.

\bibitem{Vidana:2024ngv}
I.~Vida\~na, V.~Mantovani Sarti, J.~Haidenbauer, D.~L. Mihaylov, and L.~Fabbietti.
\newblock {Neutron Star Properties and Femtoscopic Constraints}.
\newblock {\em Eur. Phys. J. A}, 61(3):59, 2025.
\newblock \href {https://arxiv.org/abs/2412.12729} {\path{arXiv:2412.12729}}, \href {https://doi.org/10.1140/epja/s10050-025-01539-z} {\path{doi:10.1140/epja/s10050-025-01539-z}}.

\bibitem{Glendenning:1991es}
N.~K. Glendenning and S.~A. Moszkowski.
\newblock {Reconciliation of neutron star masses and binding of the lambda in hypernuclei}.
\newblock {\em Phys. Rev. Lett.}, 67:2414--2417, 1991.
\newblock \href {https://doi.org/10.1103/PhysRevLett.67.2414} {\path{doi:10.1103/PhysRevLett.67.2414}}.

\bibitem{Knorren:1995ds}
R.~Knorren, M.~Prakash, and P.~J. Ellis.
\newblock {Strangeness in hadronic stellar matter}.
\newblock {\em Phys. Rev. C}, 52:3470--3482, 1995.
\newblock \href {https://arxiv.org/abs/nucl-th/9506016} {\path{arXiv:nucl-th/9506016}}, \href {https://doi.org/10.1103/PhysRevC.52.3470} {\path{doi:10.1103/PhysRevC.52.3470}}.

\bibitem{Balberg:1997yw}
Shmuel Balberg and Avraham Gal.
\newblock {An Effective equation of state for dense matter with strangeness}.
\newblock {\em Nucl. Phys. A}, 625:435--472, 1997.
\newblock \href {https://arxiv.org/abs/nucl-th/9704013} {\path{arXiv:nucl-th/9704013}}, \href {https://doi.org/10.1016/S0375-9474(97)81465-0} {\path{doi:10.1016/S0375-9474(97)81465-0}}.

\bibitem{Nishizaki:2002ih}
S.~Nishizaki, T.~Takatsuka, and Y.~Yamamoto.
\newblock {Hyperon-mixed neutron star matter and neutron stars}.
\newblock {\em Prog. Theor. Phys.}, 108:703--718, 2002.
\newblock \href {https://doi.org/10.1143/PTP.108.703} {\path{doi:10.1143/PTP.108.703}}.

\bibitem{Weissenborn:2011ut}
Simon Weissenborn, Debarati Chatterjee, and Juergen Schaffner-Bielich.
\newblock {Hyperons and massive neutron stars: vector repulsion and SU(3) symmetry}.
\newblock {\em Phys. Rev. C}, 85(6):065802, 2012.
\newblock [Erratum: Phys.Rev.C 90, 019904 (2014)].
\newblock \href {https://arxiv.org/abs/1112.0234} {\path{arXiv:1112.0234}}, \href {https://doi.org/10.1103/PhysRevC.85.065802} {\path{doi:10.1103/PhysRevC.85.065802}}.

\bibitem{Togashi:2016fky}
H.~Togashi, E.~Hiyama, Y.~Yamamoto, and M.~Takano.
\newblock {Equation of state for neutron stars with hyperons by the variational method}.
\newblock {\em Phys. Rev. C}, 93(3):035808, 2016.
\newblock \href {https://arxiv.org/abs/1602.08106} {\path{arXiv:1602.08106}}, \href {https://doi.org/10.1103/PhysRevC.93.035808} {\path{doi:10.1103/PhysRevC.93.035808}}.

\bibitem{Fortin:2017cvt}
M.~Fortin, S.~S. Avancini, C.~Provid{\^e}ncia, and I.~Vida{\~n}a.
\newblock {Hypernuclei and massive neutron stars}.
\newblock {\em Phys. Rev. C}, 95(6):065803, 2017.
\newblock \href {https://arxiv.org/abs/1701.06373} {\path{arXiv:1701.06373}}, \href {https://doi.org/10.1103/PhysRevC.95.065803} {\path{doi:10.1103/PhysRevC.95.065803}}.

\bibitem{Tong:2024jvs}
Hui Tong, Serdar Elhatisari, and Ulf-G. Mei{\ss}ner.
\newblock {Ab initio calculation of hyper-neutron matter}.
\newblock {\em Sci. Bull.}, 70:825--828, 2025.
\newblock \href {https://arxiv.org/abs/2405.01887} {\path{arXiv:2405.01887}}, \href {https://doi.org/10.1016/j.scib.2025.01.008} {\path{doi:10.1016/j.scib.2025.01.008}}.

\bibitem{Fujimoto:2024doc}
Yuki Fujimoto, Toru Kojo, and Larry McLerran.
\newblock {Quarkyonic matter pieces together the hyperon puzzle}.
\newblock 10 2024.
\newblock \href {https://arxiv.org/abs/2410.22758} {\path{arXiv:2410.22758}}.

\bibitem{Tolos:2020aln}
L.~Tolos and L.~Fabbietti.
\newblock {Strangeness in Nuclei and Neutron Stars}.
\newblock {\em Prog. Part. Nucl. Phys.}, 112:103770, 2020.
\newblock \href {https://arxiv.org/abs/2002.09223} {\path{arXiv:2002.09223}}, \href {https://doi.org/10.1016/j.ppnp.2020.103770} {\path{doi:10.1016/j.ppnp.2020.103770}}.

\bibitem{Burgio:2021vgk}
G.~F. Burgio, H.~J. Schulze, I.~Vida\~na, and J.~B. Wei.
\newblock {Neutron stars and the nuclear equation of state}.
\newblock {\em Prog. Part. Nucl. Phys.}, 120:103879, 2021.
\newblock \href {https://arxiv.org/abs/2105.03747} {\path{arXiv:2105.03747}}, \href {https://doi.org/10.1016/j.ppnp.2021.103879} {\path{doi:10.1016/j.ppnp.2021.103879}}.

\bibitem{SchaffnerBielichBook}
J\"urgen Schaffner-Bielich.
\newblock {\em {Compact Star Physics}}.
\newblock Cambridge University Press, 8 2020.
\newblock \href {https://doi.org/10.1017/9781316848357} {\path{doi:10.1017/9781316848357}}.

\bibitem{Vidana:2022tlx}
Isaac Vida{\~n}a.
\newblock {Neutron stars and the hyperon puzzle}.
\newblock {\em EPJ Web Conf.}, 271:09001, 2022.
\newblock \href {https://doi.org/10.1051/epjconf/202227109001} {\path{doi:10.1051/epjconf/202227109001}}.

\bibitem{Brueckner:1954zz}
K.~A. Brueckner, C.~A. Levinson, and H.~M. Mahmoud.
\newblock {Two-Body Forces and Nuclear Saturation. 1. Central Forces}.
\newblock {\em Phys. Rev.}, 95:217--228, 1954.
\newblock \href {https://doi.org/10.1103/PhysRev.95.217} {\path{doi:10.1103/PhysRev.95.217}}.

\bibitem{Brueckner:1955zzb}
K.~A. Brueckner and C.~A. Levinson.
\newblock {Approximate Reduction of the Many-Body Problem for Strongly Interacting Particles to a Problem of Self-Consistent Fields}.
\newblock {\em Phys. Rev.}, 97:1344--1352, 1955.
\newblock \href {https://doi.org/10.1103/PhysRev.97.1344} {\path{doi:10.1103/PhysRev.97.1344}}.

\bibitem{Day:1967zza}
B.~D. Day.
\newblock {Elements of the Brueckner-Goldstone Theory of Nuclear Matter}.
\newblock {\em Rev. Mod. Phys.}, 39:719--744, 1967.
\newblock \href {https://doi.org/10.1103/RevModPhys.39.719} {\path{doi:10.1103/RevModPhys.39.719}}.

\bibitem{Haidenbauer:2013oca}
J.~Haidenbauer, S.~Petschauer, N.~Kaiser, U.-G. Mei{\ss}ner, A.~Nogga, and W.~Weise.
\newblock {Hyperon-nucleon interaction at next-to-leading order in chiral effective field theory}.
\newblock {\em Nucl. Phys. A}, 915:24--58, 2013.
\newblock \href {https://arxiv.org/abs/1304.5339} {\path{arXiv:1304.5339}}, \href {https://doi.org/10.1016/j.nuclphysa.2013.06.008} {\path{doi:10.1016/j.nuclphysa.2013.06.008}}.

\bibitem{Haidenbauer:2019boi}
J.~Haidenbauer, U.-G. Mei{\ss}ner, and A.~Nogga.
\newblock {Hyperon--nucleon interaction within chiral effective field theory revisited}.
\newblock {\em Eur. Phys. J. A}, 56(3):91, 2020.
\newblock \href {https://arxiv.org/abs/1906.11681} {\path{arXiv:1906.11681}}, \href {https://doi.org/10.1140/epja/s10050-020-00100-4} {\path{doi:10.1140/epja/s10050-020-00100-4}}.

\bibitem{Reinert:2017usi}
P.~Reinert, H.~Krebs, and E.~Epelbaum.
\newblock {Semilocal momentum-space regularized chiral two-nucleon potentials up to fifth order}.
\newblock {\em Eur. Phys. J.}, A54(5):86, 2018.
\newblock \href {https://arxiv.org/abs/1711.08821} {\path{arXiv:1711.08821}}, \href {https://doi.org/10.1140/epja/i2018-12516-4} {\path{doi:10.1140/epja/i2018-12516-4}}.

\bibitem{J-PARCE40:2021bgw}
K.~Miwa et~al.
\newblock {Precise measurement of differential cross sections of the $\Sigma^-p \to \Lambda n$ reaction in momentum range 470-650 MeV$/c$}.
\newblock {\em Phys. Rev. Lett.}, 128(7):072501, 2022.
\newblock \href {https://arxiv.org/abs/2111.14277} {\path{arXiv:2111.14277}}, \href {https://doi.org/10.1103/PhysRevLett.128.072501} {\path{doi:10.1103/PhysRevLett.128.072501}}.

\bibitem{J-PARCE40:2021qxa}
K.~Miwa et~al.
\newblock {Measurement of the differential cross sections of the $\Sigma^-p$ elastic scattering in momentum range 470 to 850 MeV/c}.
\newblock {\em Phys. Rev. C}, 104(4):045204, 2021.
\newblock \href {https://arxiv.org/abs/2104.13608} {\path{arXiv:2104.13608}}, \href {https://doi.org/10.1103/PhysRevC.104.045204} {\path{doi:10.1103/PhysRevC.104.045204}}.

\bibitem{J-PARCE40:2022nvq}
T.~Nanamura et~al.
\newblock {Measurement of differential cross sections~for \ensuremath{\Sigma}+p elastic scattering in the momentum range 0.44\textendash{}0.80\,GeV/c}.
\newblock {\em PTEP}, 2022(9):093D01, 2022.
\newblock \href {https://arxiv.org/abs/2203.08393} {\path{arXiv:2203.08393}}, \href {https://doi.org/10.1093/ptep/ptac101} {\path{doi:10.1093/ptep/ptac101}}.

\bibitem{Le:2024rkd}
Hoai Le, Johann Haidenbauer, Ulf-G. Mei\ss{}ner, and Andreas Nogga.
\newblock {Light \ensuremath{\Lambda} Hypernuclei Studied with Chiral Hyperon-Nucleon and Hyperon-Nucleon-Nucleon Forces}.
\newblock {\em Phys. Rev. Lett.}, 134(7):072502, 2025.
\newblock \href {https://arxiv.org/abs/2409.18577} {\path{arXiv:2409.18577}}, \href {https://doi.org/10.1103/PhysRevLett.134.072502} {\path{doi:10.1103/PhysRevLett.134.072502}}.

\bibitem{Hu:2016nkw}
Jinniu Hu, Ying Zhang, Evgeny Epelbaum, Ulf-G Mei\ss{}ner, and Jie Meng.
\newblock {Nuclear matter properties with nucleon-nucleon forces up to fifth order in the chiral expansion}.
\newblock {\em Phys. Rev. C}, 96(3):034307, 2017.
\newblock \href {https://arxiv.org/abs/1612.05433} {\path{arXiv:1612.05433}}, \href {https://doi.org/10.1103/PhysRevC.96.034307} {\path{doi:10.1103/PhysRevC.96.034307}}.

\bibitem{Hu:2019zwa}
Jinniu Hu, Peiyu Wei, and Ying Zhang.
\newblock {Bayesian truncation errors in equations of state of nuclear matter with chiral nucleon-nucleon potentials}.
\newblock {\em Phys. Lett. B}, 798:134982, 2019.
\newblock \href {https://arxiv.org/abs/1909.11826} {\path{arXiv:1909.11826}}, \href {https://doi.org/10.1016/j.physletb.2019.134982} {\path{doi:10.1016/j.physletb.2019.134982}}.

\bibitem{Haidenbauer:2014uua}
J.~Haidenbauer and Ulf-G. Mei\ss{}ner.
\newblock {A study of hyperons in nuclear matter based on chiral effective field theory}.
\newblock {\em Nucl. Phys. A}, 936:29--44, 2015.
\newblock \href {https://arxiv.org/abs/1411.3114} {\path{arXiv:1411.3114}}, \href {https://doi.org/10.1016/j.nuclphysa.2015.01.005} {\path{doi:10.1016/j.nuclphysa.2015.01.005}}.

\bibitem{Petschauer:2015nea}
S.~Petschauer, J.~Haidenbauer, N.~Kaiser, Ulf-G. Mei\ss{}ner, and W.~Weise.
\newblock {Hyperons in nuclear matter from SU(3) chiral effective field theory}.
\newblock {\em Eur. Phys. J. A}, 52(1):15, 2016.
\newblock \href {https://arxiv.org/abs/1507.08808} {\path{arXiv:1507.08808}}, \href {https://doi.org/10.1140/epja/i2016-16015-4} {\path{doi:10.1140/epja/i2016-16015-4}}.

\bibitem{Mihaylov:2023ahn}
D.~L. Mihaylov, J.~Haidenbauer, and V.~Mantovani Sarti.
\newblock {Constraining the p\ensuremath{\Lambda} interaction from a combined analysis of scattering data and correlation functions}.
\newblock {\em Phys. Lett. B}, 850:138550, 2024.
\newblock \href {https://arxiv.org/abs/2312.16970} {\path{arXiv:2312.16970}}, \href {https://doi.org/10.1016/j.physletb.2024.138550} {\path{doi:10.1016/j.physletb.2024.138550}}.

\bibitem{Zheng:2025sol}
Ru-You Zheng, Zhi-Wei Liu, Li-Sheng Geng, Jin-Niu Hu, and Sibo Wang.
\newblock {In-medium \ensuremath{\Lambda}N interactions with leading order covariant chiral hyperon/nucleon-nucleon forces}.
\newblock {\em Phys. Lett. B}, 864:139416, 2025.
\newblock \href {https://arxiv.org/abs/2501.02826} {\path{arXiv:2501.02826}}, \href {https://doi.org/10.1016/j.physletb.2025.139416} {\path{doi:10.1016/j.physletb.2025.139416}}.

\bibitem{Sechi-Zorn:pLambda}
B.~Sechi-Zorn, B.~Kehoe, J.~Twitty, and R.~A. Burnstein.
\newblock {Low-Energy $\Lambda$--Proton Elastic Scattering}.
\newblock {\em Phys. Rev.}, 175:1735--1740, 1968.
\newblock \href {https://doi.org/10.1103/PhysRev.175.1735} {\path{doi:10.1103/PhysRev.175.1735}}.

\bibitem{Alexander:pLambda}
G.~Alexander, U.~Karshon, A.~Shapira, G.~Yekutieli, R.~Engelmann, H.~Filthuth, and W.~Lughofer.
\newblock {Study of the $\Lambda$--N system in low-energy $\Lambda$--p elastic scattering}.
\newblock {\em Phys. Rev.}, 173:1452--1460, 1968.
\newblock \href {https://doi.org/10.1103/PhysRev.173.1452} {\path{doi:10.1103/PhysRev.173.1452}}.

\bibitem{Epelbaum:2014efa}
E.~Epelbaum, H.~Krebs, and U.-G. Mei{\ss}ner.
\newblock {Improved chiral nucleon-nucleon potential up to next-to-next-to-next-to-leading order}.
\newblock {\em Eur. Phys. J.}, A51(5):53, 2015.
\newblock \href {https://arxiv.org/abs/1412.0142} {\path{arXiv:1412.0142}}, \href {https://doi.org/10.1140/epja/i2015-15053-8} {\path{doi:10.1140/epja/i2015-15053-8}}.

\bibitem{Epelbaum:2014sza}
E.~Epelbaum, H.~Krebs, and U.-G. Mei{\ss}ner.
\newblock {Precision nucleon-nucleon potential at fifth order in the chiral expansion}.
\newblock {\em Phys. Rev. Lett.}, 115(12):122301, 2015.
\newblock \href {https://arxiv.org/abs/1412.4623} {\path{arXiv:1412.4623}}, \href {https://doi.org/10.1103/PhysRevLett.115.122301} {\path{doi:10.1103/PhysRevLett.115.122301}}.

\bibitem{Kohno:2018gby}
M.~Kohno.
\newblock {Single-particle potential of the \ensuremath{\Lambda} hyperon in nuclear matter with chiral effective field theory NLO interactions including effects of YNN three-baryon interactions}.
\newblock {\em Phys. Rev. C}, 97(3):035206, 2018.
\newblock \href {https://arxiv.org/abs/1802.05388} {\path{arXiv:1802.05388}}, \href {https://doi.org/10.1103/PhysRevC.97.035206} {\path{doi:10.1103/PhysRevC.97.035206}}.

\bibitem{Polinder:2006eq}
H.~Polinder, J.~Haidenbauer, and U.-G. Mei{\ss}ner.
\newblock {Hyperon nucleon interactions: A chiral effective field theory approach}.
\newblock {\em Nucl. Phys. A}, 779:244--266, 2006.
\newblock URL: \url{http://dx.doi.org/10.1016/j.nuclphysa.2006.09.006}, \href {https://doi.org/10.1016/j.nuclphysa.2006.09.006} {\path{doi:10.1016/j.nuclphysa.2006.09.006}}.

\bibitem{Petschauer:2013uua}
S.~Petschauer and N.~Kaiser.
\newblock {Relativistic SU(3) chiral baryon-baryon Lagrangian up to order $q^{2}$}.
\newblock {\em Nucl. Phys.}, A916:1--29, 2013.
\newblock \href {https://arxiv.org/abs/1305.3427} {\path{arXiv:1305.3427}}.

\bibitem{Epelbaum:2004fk}
E.~Epelbaum, W.~Gl{\"o}ckle, and U.-G. Mei{\ss}ner.
\newblock {The Two-nucleon system at next-to-next-to-next-to-leading order}.
\newblock {\em Nucl. Phys.}, A747:362--424, 2005.
\newblock \href {https://arxiv.org/abs/nucl-th/0405048} {\path{arXiv:nucl-th/0405048}}.

\bibitem{Haidenbauer:2025zrr}
Johann Haidenbauer, Ulf-G. Mei{\ss}ner, and Andreas Nogga.
\newblock {Ab initio description of hypernuclei}.
\newblock 8 2025.
\newblock \href {https://arxiv.org/abs/2508.05243} {\path{arXiv:2508.05243}}.

\bibitem{HypernuclearDataBase}
P.~Eckert, P.~Achenbach, et~al.
\newblock Chart of hypernuclides --- {H}ypernuclear structure and decay data, 2021.
\newblock \href{https://hypernuclei.kph.uni-mainz.de}{https://hypernuclei.kph.uni-mainz.de}.

\bibitem{Petschauer:2015elq}
Stefan Petschauer, Norbert Kaiser, Johann Haidenbauer, Ulf-G. Mei{\ss}ner, and Wolfram Weise.
\newblock {Leading three-baryon forces from SU(3) chiral effective field theory}.
\newblock {\em Phys. Rev. C}, 93(1):014001, 2016.
\newblock \href {https://arxiv.org/abs/1511.02095} {\path{arXiv:1511.02095}}, \href {https://doi.org/10.1103/PhysRevC.93.014001} {\path{doi:10.1103/PhysRevC.93.014001}}.

\bibitem{Reuber:1993ip}
A.~Reuber, K.~Holinde, and J.~Speth.
\newblock {Meson exchange hyperon - nucleon interactions in free scattering and nuclear matter}.
\newblock {\em Nucl. Phys. A}, 570:543--579, 1994.
\newblock \href {https://doi.org/10.1016/0375-9474(94)90073-6} {\path{doi:10.1016/0375-9474(94)90073-6}}.

\bibitem{Rijken:1998yy}
T.~A. Rijken, V.~G.~J. Stoks, and Y.~Yamamoto.
\newblock {Soft core hyperon - nucleon potentials}.
\newblock {\em Phys. Rev. C}, 59:21--40, 1999.
\newblock \href {https://arxiv.org/abs/nucl-th/9807082} {\path{arXiv:nucl-th/9807082}}, \href {https://doi.org/10.1103/PhysRevC.59.21} {\path{doi:10.1103/PhysRevC.59.21}}.

\bibitem{Kohno:1999nz}
M.~Kohno, Y.~Fujiwara, T.~Fujita, C.~Nakamoto, and Y.~Suzuki.
\newblock {Hyperon single particle potentials calculated from SU(6) quark model baryon baryon interactions}.
\newblock {\em Nucl. Phys. A}, 674:229--245, 2000.
\newblock \href {https://arxiv.org/abs/nucl-th/9912059} {\path{arXiv:nucl-th/9912059}}, \href {https://doi.org/10.1016/S0375-9474(00)00164-0} {\path{doi:10.1016/S0375-9474(00)00164-0}}.

\bibitem{Schulze:1998jf}
H.~J. Schulze, M.~Baldo, U.~Lombardo, J.~Cugnon, and A.~Lejeune.
\newblock {Hyperonic nuclear matter in Bruckner theory}.
\newblock {\em Phys. Rev. C}, 57:704--713, 1998.
\newblock \href {https://doi.org/10.1103/PhysRevC.57.704} {\path{doi:10.1103/PhysRevC.57.704}}.

\bibitem{Vidana:1999jm}
I.~Vida\~na, A.~Polls, A.~Ramos, M.~Hjorth-Jensen, and V.~G.~J. Stoks.
\newblock {Strange nuclear matter within Bruckner-Hartree-Fock theory}.
\newblock {\em Phys. Rev. C}, 61:025802, 2000.
\newblock \href {https://arxiv.org/abs/nucl-th/9909019} {\path{arXiv:nucl-th/9909019}}, \href {https://doi.org/10.1103/PhysRevC.61.025802} {\path{doi:10.1103/PhysRevC.61.025802}}.

\bibitem{Song:1998zz}
H.~Q. Song, M.~Baldo, G.~Giansiracusa, and U.~Lombardo.
\newblock {Bethe-Brueckner-Goldstone Expansion in Nuclear Matter}.
\newblock {\em Phys. Rev. Lett.}, 81:1584--1587, 1998.
\newblock \href {https://doi.org/10.1103/PhysRevLett.81.1584} {\path{doi:10.1103/PhysRevLett.81.1584}}.

\bibitem{Baldo:2001mv}
M.~Baldo, A.~Fiasconaro, H.~Q. Song, G.~Giansiracusa, and U.~Lombardo.
\newblock {High density symmetric nuclear matter in the Bethe-Brueckner-Goldstone approach}.
\newblock {\em Phys. Rev. C}, 65:017303, 2002.
\newblock \href {https://doi.org/10.1103/PhysRevC.65.017303} {\path{doi:10.1103/PhysRevC.65.017303}}.

\bibitem{Lu:2017nbi}
Jia-Jing Lu, Zeng-Hua Li, Chong-Yang Chen, M.~Baldo, and H.-J. Schulze.
\newblock {Convergence of the hole-line expansion with modern nucleon-nucleon potentials}.
\newblock {\em Phys. Rev. C}, 96(4):044309, 2017.
\newblock \href {https://doi.org/10.1103/PhysRevC.96.044309} {\path{doi:10.1103/PhysRevC.96.044309}}.

\bibitem{Kohno:2013ihv}
M.~Kohno.
\newblock {Nuclear and neutron matter G-matrix calculations with a chiral effective field theory potential including effects of three-nucleon interactions}.
\newblock {\em Phys. Rev. C}, 88(6):064005, 2013.
\newblock \href {https://arxiv.org/abs/1309.4556} {\path{arXiv:1309.4556}}, \href {https://doi.org/10.1103/PhysRevC.88.064005} {\path{doi:10.1103/PhysRevC.88.064005}}.

\bibitem{Gerstung:2020fzk}
Dominik Gerstung.
\newblock {Hyperons in nuclear matter and SU(3) chiral effective field theory, PhD thesis}, 2020.

\bibitem{Entem:2003ft}
D.R. Entem and R.~Machleidt.
\newblock {Accurate charge dependent nucleon nucleon potential at fourth order of chiral perturbation theory}.
\newblock {\em Phys. Rev. C}, 68:041001, 2003.
\newblock \href {https://arxiv.org/abs/nucl-th/0304018} {\path{arXiv:nucl-th/0304018}}, \href {https://doi.org/10.1103/PhysRevC.68.041001} {\path{doi:10.1103/PhysRevC.68.041001}}.

\bibitem{Hu:private}
Jinniu Hu.
\newblock {private communication}.

\bibitem{Holt:2009uk}
J.~W. Holt, N.~Kaiser, and W.~Weise.
\newblock {Chiral three-nucleon interaction and the C-14 dating beta decay}.
\newblock {\em Phys. Rev. C}, 79:054331, 2009.
\newblock \href {https://arxiv.org/abs/0901.4750} {\path{arXiv:0901.4750}}, \href {https://doi.org/10.1103/PhysRevC.79.054331} {\path{doi:10.1103/PhysRevC.79.054331}}.

\bibitem{Holt:2009ty}
J.~W. Holt, N.~Kaiser, and W.~Weise.
\newblock {Density-dependent effective nucleon-nucleon interaction from chiral three-nucleon forces}.
\newblock {\em Phys. Rev. C}, 81:024002, 2010.
\newblock \href {https://arxiv.org/abs/0910.1249} {\path{arXiv:0910.1249}}, \href {https://doi.org/10.1103/PhysRevC.81.024002} {\path{doi:10.1103/PhysRevC.81.024002}}.

\bibitem{Nagels:2015lfa}
M.~M. Nagels, Th.~A. Rijken, and Y.~Yamamoto.
\newblock {Extended-soft-core baryon-baryon model ESC16. II. Hyperon-nucleon interactions}.
\newblock {\em Phys. Rev. C}, 99(4):044003, 2019.
\newblock \href {https://arxiv.org/abs/1501.06636} {\path{arXiv:1501.06636}}, \href {https://doi.org/10.1103/PhysRevC.99.044003} {\path{doi:10.1103/PhysRevC.99.044003}}.

\bibitem{Gal:2016boi}
A.~Gal, E.V. Hungerford, and D.J. Millener.
\newblock {Strangeness in nuclear physics}.
\newblock {\em Rev. Mod. Phys.}, 88(3):035004, 2016.
\newblock \href {https://arxiv.org/abs/1605.00557} {\path{arXiv:1605.00557}}, \href {https://doi.org/10.1103/RevModPhys.88.035004} {\path{doi:10.1103/RevModPhys.88.035004}}.

\bibitem{Friedman:2023ucs}
E.~Friedman and A.~Gal.
\newblock {\ensuremath{\Lambda} hypernuclear potentials beyond linear density dependence}.
\newblock {\em Nucl. Phys. A}, 1039:122725, 2023.
\newblock \href {https://arxiv.org/abs/2306.06973} {\path{arXiv:2306.06973}}, \href {https://doi.org/10.1016/j.nuclphysa.2023.122725} {\path{doi:10.1016/j.nuclphysa.2023.122725}}.

\bibitem{Krebs:2023gge}
Hermann Krebs and Evgeny Epelbaum.
\newblock {Toward consistent nuclear interactions from chiral Lagrangians. II. Symmetry preserving regularization}.
\newblock {\em Phys. Rev. C}, 110(4):044004, 2024.
\newblock \href {https://arxiv.org/abs/2312.13932} {\path{arXiv:2312.13932}}, \href {https://doi.org/10.1103/PhysRevC.110.044004} {\path{doi:10.1103/PhysRevC.110.044004}}.

\bibitem{Wiringa:1994wb}
Robert~B. Wiringa, V.G.J. Stoks, and R.~Schiavilla.
\newblock {An Accurate nucleon-nucleon potential with charge independence breaking}.
\newblock {\em Phys. Rev. C}, 51:38, 1995.
\newblock \href {https://arxiv.org/abs/nucl-th/9408016} {\path{arXiv:nucl-th/9408016}}, \href {https://doi.org/10.1103/PhysRevC.51.38} {\path{doi:10.1103/PhysRevC.51.38}}.

\bibitem{Isaule:2016pnn}
Felipe Isaule, H.~F. Arellano, and Arnau Rios.
\newblock {Di-neutrons in neutron matter within a Brueckner-Hartree-Fock approach}.
\newblock {\em Phys. Rev. C}, 94(3):034004, 2016.
\newblock \href {https://arxiv.org/abs/1602.05234} {\path{arXiv:1602.05234}}, \href {https://doi.org/10.1103/PhysRevC.94.034004} {\path{doi:10.1103/PhysRevC.94.034004}}.

\bibitem{Chorozidou:2024gyy}
Arsenia Chorozidou and Theodoros Gaitanos.
\newblock {Momentum dependence of in-medium potentials: A solution to the hyperon puzzle in neutron stars}.
\newblock {\em Phys. Rev. C}, 109(3):L032801, 2024.
\newblock \href {https://doi.org/10.1103/PhysRevC.109.L032801} {\path{doi:10.1103/PhysRevC.109.L032801}}.

\bibitem{Nara:2022kbb}
Yasushi Nara, Asanosuke Jinno, Koichi Murase, and Akira Ohnishi.
\newblock {Directed flow of {\ensuremath{\Lambda}} in high-energy heavy-ion collisions and {\ensuremath{\Lambda}} potential in dense nuclear matter}.
\newblock {\em Phys. Rev. C}, 106(4):044902, 2022.
\newblock \href {https://arxiv.org/abs/2208.01297} {\path{arXiv:2208.01297}}, \href {https://doi.org/10.1103/PhysRevC.106.044902} {\path{doi:10.1103/PhysRevC.106.044902}}.

\bibitem{Yamamoto:1988qz}
Y.~Yamamoto, H.~Bando, and J.~Zofka.
\newblock {On the $\Lambda$ Hypernuclear Single Particle Energies}.
\newblock {\em Prog. Theor. Phys.}, 80:757--761, 1988.
\newblock \href {https://doi.org/10.1143/PTP.80.757} {\path{doi:10.1143/PTP.80.757}}.

\bibitem{Jinno:2023xjr}
Asanosuke Jinno, Koichi Murase, Yasushi Nara, and Akira Ohnishi.
\newblock {Repulsive {\ensuremath{\Lambda}} potentials in dense neutron star matter and binding energy of {\ensuremath{\Lambda}} in hypernuclei}.
\newblock {\em Phys. Rev. C}, 108(6):065803, 2023.
\newblock \href {https://arxiv.org/abs/2306.17452} {\path{arXiv:2306.17452}}, \href {https://doi.org/10.1103/PhysRevC.108.065803} {\path{doi:10.1103/PhysRevC.108.065803}}.

\bibitem{Nogami:1970rk}
Y.~Nogami and E.~Satoh.
\newblock {Effect of lambda sigma conversion on the lambda-particle binding in nuclear matter}.
\newblock {\em Nucl. Phys. B}, 19:93--106, 1970.
\newblock \href {https://doi.org/10.1016/0550-3213(70)90031-3} {\path{doi:10.1016/0550-3213(70)90031-3}}.

\bibitem{Bodmer:1971eh}
A.~R. Bodmer and D.~M. Rote.
\newblock {Lambda-n sigma-n coupling for lambda-n scattering and for the lambda-particle binding in nuclear matter}.
\newblock {\em Nucl. Phys. A}, 169:1--48, 1971.
\newblock \href {https://doi.org/10.1016/0375-9474(71)90556-2} {\path{doi:10.1016/0375-9474(71)90556-2}}.

\bibitem{Dabrowski:1973enk}
J.~Dabrowski.
\newblock {On the effect of lambda sigma conversion on the lambda particle binding energy in nuclear matter}.
\newblock {\em Phys. Lett. B}, 47:306--310, 1973.
\newblock \href {https://doi.org/10.1016/0370-2693(73)90609-6} {\path{doi:10.1016/0370-2693(73)90609-6}}.

\bibitem{Fujiwara:2006yh}
Y.~Fujiwara, Y.~Suzuki, and C.~Nakamoto.
\newblock {Baryon-baryon interactions in the SU(6) quark model and their applications to light nuclear systems}.
\newblock {\em Prog. Part. Nucl. Phys.}, 58:439--520, 2007.
\newblock \href {https://arxiv.org/abs/nucl-th/0607013} {\path{arXiv:nucl-th/0607013}}, \href {https://doi.org/10.1016/j.ppnp.2006.08.001} {\path{doi:10.1016/j.ppnp.2006.08.001}}.

\bibitem{ALICE:2025plu}
Ibrahim~Jaser Abualrob et~al.
\newblock {Measurement of the p-$\Sigma^+$ correlation function in pp collisions at $\sqrt{\textit{s}}=13$ TeV}.
\newblock 10 2025.
\newblock \href {https://arxiv.org/abs/2510.14448} {\path{arXiv:2510.14448}}.

\bibitem{Jinno:2025mos}
Asanosuke Jinno, Koichi Murase, and Yasushi Nara.
\newblock {{\ensuremath{\Lambda}} and {\ensuremath{\Sigma}} potentials in neutron stars, hypernuclei, and heavy-ion collisions}.
\newblock {\em PoS}, EXA-LEAP2024:034, 2025.
\newblock \href {https://arxiv.org/abs/2501.09881} {\path{arXiv:2501.09881}}, \href {https://doi.org/10.22323/1.480.0034} {\path{doi:10.22323/1.480.0034}}.

\bibitem{Mares:1995bm}
J.~Mares, E.~Friedman, A.~Gal, and B.~K. Jennings.
\newblock {Constraints on Sigma nucleus dynamics from Dirac phenomenology of Sigma- atoms}.
\newblock {\em Nucl. Phys. A}, 594:311--324, 1995.
\newblock \href {https://arxiv.org/abs/nucl-th/9505003} {\path{arXiv:nucl-th/9505003}}, \href {https://doi.org/10.1016/0375-9474(95)00358-8} {\path{doi:10.1016/0375-9474(95)00358-8}}.

\bibitem{Dabrowski:1999jy}
J.~Dabrowski.
\newblock {Isospin dependence of the single particle potential of the Sigma hyperon in nuclear matter}.
\newblock {\em Phys. Rev. C}, 60:025205, 1999.
\newblock \href {https://doi.org/10.1103/PhysRevC.60.025205} {\path{doi:10.1103/PhysRevC.60.025205}}.

\bibitem{Harada:2006yj}
T.~Harada and Y.~Hirabayashi.
\newblock {Sigma- production spectrum in the inclusive (pi-, K+) reaction on Bi-209 and the Sigma-nucleus potential}.
\newblock {\em Nucl. Phys. A}, 767:206--217, 2006.
\newblock \href {https://doi.org/10.1016/j.nuclphysa.2005.12.018} {\path{doi:10.1016/j.nuclphysa.2005.12.018}}.

\bibitem{Kohno:2006iq}
M.~Kohno, Y.~Fujiwara, Y.~Watanabe, K.~Ogata, and M.~Kawai.
\newblock {Semiclassical distorted wave model analysis of the (pi-,K+) Sigma formation inclusive spectrum}.
\newblock {\em Phys. Rev. C}, 74:064613, 2006.
\newblock \href {https://arxiv.org/abs/nucl-th/0611080} {\path{arXiv:nucl-th/0611080}}, \href {https://doi.org/10.1103/PhysRevC.74.064613} {\path{doi:10.1103/PhysRevC.74.064613}}.

\bibitem{Harada:2023otu}
Toru Harada and Yoshiharu Hirabayashi.
\newblock {Production spectra with a {\ensuremath{\Sigma}}{\ensuremath{-}} hyperon in ({\ensuremath{\pi}}{\ensuremath{-}},K+) reactions on light to heavy nuclei}.
\newblock {\em Phys. Rev. C}, 107(5):054611, 2023.
\newblock \href {https://doi.org/10.1103/PhysRevC.107.054611} {\path{doi:10.1103/PhysRevC.107.054611}}.

\bibitem{Dover:1984nr}
C.~B. Dover, A.~Gal, and D.~J. Millener.
\newblock {IS ISOSPIN A GOOD QUANTUM NUMBER FOR SIGMA HYPERNUCLEI?}
\newblock {\em Phys. Lett. B}, 138:337--340, 1984.
\newblock \href {https://doi.org/10.1016/0370-2693(84)91911-7} {\path{doi:10.1016/0370-2693(84)91911-7}}.

\bibitem{Kohno:2009vk}
M.~Kohno and Y.~Fujiwara.
\newblock {Localized N, Lambda, Sigma, and Xi Single-Particle Potentials in Finite Nuclei Calculated with SU(6) Quark-Model Baryon-Baryon Interactions}.
\newblock {\em Phys. Rev. C}, 79:054318, 2009.
\newblock \href {https://arxiv.org/abs/0904.0517} {\path{arXiv:0904.0517}}, \href {https://doi.org/10.1103/PhysRevC.79.054318} {\path{doi:10.1103/PhysRevC.79.054318}}.

\bibitem{LENPIC:2015qsz}
S.~Binder et~al.
\newblock {Few-nucleon systems with state-of-the-art chiral nucleon-nucleon forces}.
\newblock {\em Phys. Rev. C}, 93(4):044002, 2016.
\newblock \href {https://arxiv.org/abs/1505.07218} {\path{arXiv:1505.07218}}, \href {https://doi.org/10.1103/PhysRevC.93.044002} {\path{doi:10.1103/PhysRevC.93.044002}}.

\bibitem{Le:2023bfj}
H.~Le, J.~Haidenbauer, U.-G. Mei\ss{}ner, and A.~Nogga.
\newblock {Separation energies of light $\Lambda$ hypernuclei and their theoretical uncertainties}.
\newblock {\em Eur. Phys. J. A}, 60(1):3, 2024.
\newblock \href {https://doi.org/10.1140/epja/s10050-023-01219-w} {\path{doi:10.1140/epja/s10050-023-01219-w}}.

\bibitem{Furnstahl:2015rha}
R.~J. Furnstahl, N.~Klco, D.~R. Phillips, and S.~Wesolowski.
\newblock {Quantifying truncation errors in effective field theory}.
\newblock {\em Phys. Rev. C}, 92(2):024005, 2015.
\newblock \href {https://arxiv.org/abs/1506.01343} {\path{arXiv:1506.01343}}, \href {https://doi.org/10.1103/PhysRevC.92.024005} {\path{doi:10.1103/PhysRevC.92.024005}}.

\bibitem{Sammarruca:2014zia}
F.~Sammarruca, L.~Coraggio, J.~W. Holt, N.~Itaco, R.~Machleidt, and L.~E. Marcucci.
\newblock {Toward order-by-order calculations of the nuclear and neutron matter equations of state in chiral effective field theory}.
\newblock {\em Phys. Rev. C}, 91(5):054311, 2015.
\newblock \href {https://arxiv.org/abs/1411.0136} {\path{arXiv:1411.0136}}, \href {https://doi.org/10.1103/PhysRevC.91.054311} {\path{doi:10.1103/PhysRevC.91.054311}}.

\bibitem{Shang:2021msd}
Xin-Le Shang, Jian-Min Dong, Wei Zuo, Peng Yin, and U.~Lombardo4.
\newblock {Exact solution of the Brueckner-Bethe-Goldstone equation with three-body forces in nuclear matter}.
\newblock {\em Phys. Rev. C}, 103(3):034316, 2021.
\newblock \href {https://arxiv.org/abs/2107.12228} {\path{arXiv:2107.12228}}, \href {https://doi.org/10.1103/PhysRevC.103.034316} {\path{doi:10.1103/PhysRevC.103.034316}}.

\bibitem{Melendez:2019izc}
J.~A. Melendez, R.~J. Furnstahl, D.~R. Phillips, M.~T. Pratola, and S.~Wesolowski.
\newblock {Quantifying Correlated Truncation Errors in Effective Field Theory}.
\newblock {\em Phys. Rev. C}, 100(4):044001, 2019.
\newblock \href {https://arxiv.org/abs/1904.10581} {\path{arXiv:1904.10581}}, \href {https://doi.org/10.1103/PhysRevC.100.044001} {\path{doi:10.1103/PhysRevC.100.044001}}.

\bibitem{Bethe:1965zz}
H.~A. Bethe.
\newblock {Three-Body Correlations in Nuclear Matter}.
\newblock {\em Phys. Rev.}, 138:B804--B822, 1965.
\newblock \href {https://doi.org/10.1103/PhysRev.138.B804} {\path{doi:10.1103/PhysRev.138.B804}}.

\bibitem{Day:1966zza}
Ben Day.
\newblock {Improved Solution to the Bethe-Faddeev Equations}.
\newblock {\em Phys. Rev.}, 151:826--829, 1966.
\newblock \href {https://doi.org/10.1103/PhysRev.151.826} {\path{doi:10.1103/PhysRev.151.826}}.

\bibitem{Sammarruca:2018va}
F.~Sammarruca, Laura~E. Marcucci, L.~Coraggio, J.~W. Holt, N.~Itaco, and R.~Machleidt.
\newblock {Nuclear and neutron matter equations of state from high-quality potentials up to fifth order of the chiral expansion}.
\newblock July 2018.
\newblock URL: \url{https://arxiv.org/abs/1807.06640}, \href {https://arxiv.org/abs/1807.06640} {\path{arXiv:1807.06640}}.

\bibitem{Holt:2019bah}
Jeremy~W. Holt, Mamiya Kawaguchi, and Norbert Kaiser.
\newblock {Implementing chiral three-body forces in terms of medium-dependent two-body forces}.
\newblock {\em Front. in Phys.}, 8:100, 2020.
\newblock \href {https://arxiv.org/abs/1912.06055} {\path{arXiv:1912.06055}}, \href {https://doi.org/10.3389/fphy.2020.00100} {\path{doi:10.3389/fphy.2020.00100}}.

\bibitem{Petschauer:2016pbn}
Stefan Petschauer, Johann Haidenbauer, Norbert Kaiser, Ulf-G. Mei\ss{}ner, and Wolfram Weise.
\newblock {Density-dependent effective baryon\textendash{}baryon interaction from chiral three-baryon forces}.
\newblock {\em Nucl. Phys. A}, 957:347--378, 2017.
\newblock \href {https://arxiv.org/abs/1607.04307} {\path{arXiv:1607.04307}}, \href {https://doi.org/10.1016/j.nuclphysa.2016.09.010} {\path{doi:10.1016/j.nuclphysa.2016.09.010}}.

\end{thebibliography}

\end{document}